\documentstyle[]{article}

\begin{document}

\title{Quantum processes on phase space}
\author{Charis Anastopoulos \thanks{anastop@phys.uu.nl}, \\
\\Spinoza Instituut, Leuvenlaan 4, \\
3584HE Utrecht, The Netherlands  } \maketitle

\begin{abstract}
Quantum theory predicts probabilities for various events as well as relative phases
(interference or geometric) between different alternatives of the system. The most
general description of the latter is in terms of the Pancharatnam phase. A unified
description of  both probabilities and phases comes through generalisation of the
notion of a density matrix for histories; this object is the decoherence functional
introduced by the consistent histories approach.  If we take phases as well as
probabilities as primitive elements of our theory, we abandon Kolmogorov probability
and can describe quantum theory
 in terms of fundamental commutative observables, without being obstructed by Bell's and
related theorems.

We develop the description of relative phases and probabilities for paths on the
classical phase space. This description provides  a theory of   quantum processes,
having many formal analogies with the theory of
 stochastic processes. We identify a number of basic postulates and study its corresponding properties.
We  strongly emphasise    the notion of conditioning (by generalising classical
conditional probability) and are able to write ``quantum differential equations'' as
analogous to stochastic differential equations. They can be interpreted as referring
to   individual systems.

We, then, show the sense by which quantum theory is equivalent to a quantum process on
phase space (using coherent states).
 Conversely starting from quantum processes on phase space we recover standard
quantum theory on Hilbert space from the requirement that the process satisfies (an
analogue of ) the Markov property together with time reversibility. The statistical
predictions of our theory are identical to the ones of standard quantum theory, but
the ``logic'' of events is Boolean;  events are not represented by projectors any
more. We discuss  some implication of this fact for the interpretation of quantum
theory, emphasising that it makes plausible the existence of realist theories for
individual quantum systems.

\end{abstract}

\renewcommand {\thesection}{\arabic{section}}
 \renewcommand {\theequation}{\thesection. \arabic{equation}}
\let \ssection = \section
\renewcommand{\section}{\setcounter{equation}{0} \ssection}

\section{Introduction}
\subsection{Quantum phases and probabilities}
Quantum mechanics is a theory of probabilities, but  it is not a probability theory,
in the standard use of this term. The reason for this lies in the fact that the basic
objects in quantum theory are vectors in a {\em complex} Hilbert space (wave
functions) and probabilities are obtained by squaring them. Hence, the basic quantum
object is  the probability amplitude, not the probability itself.

When we deal with properties of systems at a single moment of time, this distinction
loses some of its immediacy. Indeed, one can describe quantum theory in terms of
states (density matrices), observables (self-adjoint operators) and propositions
(projection operators): these concepts are identical to the ones employed by classical
probability theory. The only difference lies in the non-commutativity of the space of
observables.

However, this exact correspondence evaporates when we consider the study of properties
of the system at more than one moments of time. In this case the underlying complex
nature of the basic quantities manifests itself in the presence of phases, either ones
associated to interference or geometric ones.

The two-slit experiment is well known as a prototype of quantum mechanical
interference. It is a special case of a general mathematical property of quantum
theory: {\em probabilities corresponding to histories are non-additive}, i.e. if we
have two histories  (propositions) $\alpha$ and $\beta$ then
\begin{equation}
p(\alpha \vee  \beta) \neq p(\alpha) + p(\beta)
\end{equation}
Here $p$ refers to the probability distribution associated to a state and $\vee$
refers to the conjunction of histories viewed as propositions.

This fact is the starting point of the consistent histories approach
\cite{Gri84,Omn8894, GeHa9093, Har93a, I94}, which is a realist interpretation of
quantum theory. It remarks that the probabilistic information about histories can be
incorporated into a function of pairs of histories, the decoherence functional.
Formally, this is  a generalisation of the notion of the density matrix in the history
context. Whenever  we have a set of histories in which  the off-diagonal elements of
the decoherence functional are vanishing, its diagonal elements can be unambiguously
interpreted as a probability distribution {\em for this set of histories}. However the
consistent histories approach does not offer a natural physical interpretation for the
off-diagonal elements of the decoherence functional.

Besides the non-additivity of probabilities the complex nature of quantum theory
manifests itself in the importance of geometric phases, i.e. phases that appear during
the quantum evolution of the system and  depend only on the path transversed and not
on the duration of the evolution or the Hamiltonian that drives the system. The
paradigmatic example is the Berry phase \cite{Ber84, Sim83, WilS89}, which was
originally derived for cyclic, adiabatic evolution. However, it was soon realised that
the geometric phase exists for non-cyclic, non-adiabatic evolutions \cite{AnaAh87}
and, what is important for our present context, for sequences of measurements
\cite{SaBha88}. An important property of the geometric phase is that it can be
measured only relatively, i.e. by comparing the evolution of two distinct ensembles of
systems through their interference pattern.

The geometric phase and the histories approach can be viewed in a unified framework by
 noticing that the geometric phase is the {\em building block of the decoherence
  functional}. In \cite{AnSav02} it was shown that for the limit of continuous-time
histories the off-diagonal elements of the decoherence functional  amount
 to the difference in geometric phase between the paths,
thus providing an interpretation
  lacking in the consistent histories approach. This interpretation persists
for the case of coarse-grained histories at discrete moments of time, by virtue of the
generalisations of the geometric phase \cite{An01a}. We show in this paper (section
2.2) that the most natural interpretation of the off-diagonal elements of the
decoherence functional is in terms of the geometric phase that arose out of the work
of Pancharatnam \cite{Pan56, SaBha88}.

The picture that emerges from these results is that a formalism for quantum theory
that
 is based on histories, needs to include not only the notion of probabilities, but also
  that of the relative phases. If we take the phases as elementary ingredients of the
   theory we abandon  the idea that quantum theory is to be based on additive probabilities.
But then the classic non-go theorems of Bell-Wigner \cite{Bell64} or Kochen-Specker
\cite{KoSp67} are not forbidding for the description of the quantum system in terms of
some version of ``hidden variables'', that
 are not deterministic. In \cite{An01a} we showed that
one can recover all statistical predictions of quantum theory (essentially the
correlation functions) from a theory that {\em is based on commutative observables},
essentially functions on the classical phase space. The only difference is in the
notion of events: in quantum theory events are said to correspond to projection
operators (and hence form a non-distributive lattice), while in our construction they
correspond to subsets of phase space and as such they form a classical Boolean
lattice.

\subsection{This paper}

An  epigrammatic way to state our previous discussion is that if we treat quantum
phases as basic ingredients of quantum theory allows us to get the predictions of
standard quantum theory, while {\em dispensing with quantum logic}.

This statement is the starting point of this paper. We want to develop a statistical
theory for histories on a  manifold (usually a symplectic manifold), which will be
based on {\em a decoherence functional rather than a probability measure}. We take as
prototype the theory of {\em stochastic processes}, and for this reason we call this
class of theories{\em quantum processes}. We need to warn the reader that this name is
not used in the same sense as has been used in the literature, where it was used to
denote stochastic processes on the quantum mechanical Hilbert space (see \cite{Str00}
for a recent review).

Two are the main ingredients of the theory of quantum processes. One first needs to
specify, what the events are. We take an elementary event  to correspond to a point of
a sample space $\Omega$ and
 a (coarse-grained) event  is a {\em measurable subset} of $\Omega$. In the context of histories $\Omega$ is taken as
a space of suitable maps from a subset of the real line ${\bf R}$ to a manifold
$\Gamma$. We shall mostly identify  $\Gamma$ with the classical phase space.

The other ingredient is the decoherence functional: it is a complex-valued functional
that takes a pair of histories to a complex number {\em in the unit disc}. Its
defining properties are essentially these of a density matrix; in fact they have a
direct operational significance, in terms of the standard procedure of measuring the
relative phase between different histories of the system. The only condition without a
direct operational significance is the additivity, which is a manifestation of  the
{\em superposition} principle of standard quantum theory. When we have histories, we
can define the decoherence functional in terms of its restriction on discrete moments
of time. In a   fashion similar to classical probability theory, a decoherence
functional on continuous-time histories can be defined by a hierarchy of multi-time
hierarchy distributions that satisfy a condition of compatibility.

Again in analogy to classical probability theory, we introduce the notion of
conditioning, i.e. the changing of the physical quantities in such a way as to take
into account the fact that we focus on a particular class of observables. Conditioning
takes place with respect to a subalgebra of observables, which correspond to the
particular class of events to which we focus our description. The standard result of
the ``reduction of the wave packet'' arises as a special case.

Now, whenever we have histories there is a sharp distinction between the notion of
time as manifested in kinematics as compared with the way it is manifested in
dynamics. This was shown by Savvidou \cite{Sav99a}, in the context of the
continuous-time  formulation of consistent histories, but makes sense in any theory
that has histories as basic objects (whether this is quantum theory or classical
mechanics or the theory of stochastic processes). In effect, the histories description
allows a distinction between the kinematical and dynamical aspects of time,which is
implemented by different symmetry groups \cite{Sav01a, Sav01b}. In quantum theory, it
is argued that this distinction manifests itself in the distinction between
``reduction of the wave packet'' and Heisenberg dynamics \cite{Sav99a} or between the
geometric and dynamical phases \cite{AnSav02}.

For this reason it is convenient to distinguish between the kinematical and dynamical
aspects of quantum processes; we distinguish the kinematic process, which is obtained
for vanishing Hamiltonian: this is unlike classical probability theory, where the
kinematic processes are trivial; in quantum theory the kinematic process contains {\em
all characteristic features of quantum theory}. We construct such processes for
standard quantum theory, by employing coherent states. The introduction of dynamics is
then rather straightforward;  one can write {\em quantum differential equations} in
analogy with stochastic differential equations, in which ``fluctuating forces''
subject to  the kinematical process  complement deterministic evolution. The kinematic
process is then a building block for quantum processes: in effect it is  the analogue
of the Wiener process in the theory of stochastic processes, which is usually employed
in order to model a fluctuating environment.

We then  proceed to argue inversely: given a quantum process on phase space,
 how do we recover standard quantum theory? The answer is very simple: if we assume that
the process satisfies an analogue of the Markov property and time-reversibility, then
this process  can be described by quantum theory on the {\em standard Hilbert space}.
Dropping time-reversibility one gets the theory of open quantum systems. It is,
however, important to make an assumption of continuity for the propagator, which
amounts to demanding the existence of a {\em non-trivial } kinematic process
associated to the system under study.

Eventually, we arrive at  the point we aimed to: a theory that has the same
statistical predictions of quantum theory, but whose notion of events is different; an
event (or a sharp measurement) is not associated to a projection operator, but to a
phase space cell. Hence the ``logic'' of these events is Boolean. In comparing with
standard quantum theory, this means that we drop the principle that measurable
quantities correspond to {\em eigenvalues} of the operator observables. We argue that
this principle has actually no compelling experimental evidence and is rather made for
historical reasons and for purposes of convenience. The discreteness manifested in
many quantum phenomena can eventually be attributed to dynamics or a combination of
dynamics and kinematics. After all, this kind  of information is contained in the
correlation functions, which are obtainable by a quantum process, without any
assumption about a special status of operators' eigenvalues.

Our analysis demonstrates, that it is possible to set theoretical frameworks that can
provide the same statistical predictions to the ones of quantum theory, while having
as observables only real-valued functions. Quantum theory is then possible without
``quantum logic''. Even though our construction is operational in character and cannot
claim to be a theory for the individual quantum system (like Kopenhagen quantum
theory), it is not incompatible with such a theory. And by virtue of the preservation
of classical logic no issues  of contextuality -like one exhibited in the
Kochen-Specker ``paradox''-  are  likely to arise.
\\ \\
Throughout the paper there are three points, the consequences of which we invariably point out: \\ \\
i. The formal analogy of quantum processes to classical stochastic processes, that
allows us to employ successful techniques from the latter theory in the former. For
this paper we have tried to keep  the perspective  on the rigorous presentation as in
\cite{DeWitt}, but we have found particularly useful the  semi-formal treatment
of \cite{HoLe84} and the perspective afforded by \cite{Str00, Str96}. \\
ii. The fact that the fundamental behaviour of quantum  theory is already present at the kinematical level. In particular, the presence of complex numbers as  manifested in the phases is not due to the Schr\"odinger equation as often assumed, but goes deeper in the fundamental set up of the theory. \\
iii. Our eventual aim is to find a description for the individual system, that would
reproduce the description in terms of quantum processes in the statistics. We have
reasons to believe that such a description might be possible to be based in the
geometry of the classical phase space, or rather on additional geometric structures
introduced in the classical phase space. For this reason, we give some emphasis in
possible geometric origins of our basic objects.

\section{Background}

\subsection{The standard histories formalism}

A history  corresponds to a sequence of projection operators $\hat{\alpha}_{t_1},
\ldots, \hat{\alpha}_{t_n}$, and it corresponds to a time-ordered sequence of
propositions about the physical system. The indices $t_1, \ldots, t_n$ {\it refer to
the time a proposition is asserted and have no dynamical meaning.} Dynamics are
related to the Hamiltonian $\hat{H}$, which defines the one-parameter group of unitary
operators $\hat{U}(s) = e^{-i\hat{H}s}$.

A natural way to represent the space of all histories is by defining a history Hilbert
space ${\cal V} := \otimes_{t_i} {\cal H}_{t_i}$, where ${\cal H}_{t_i}$ is a copy of
the standard Hilbert space, indexed by the moment of time to which it corresponds. A
history is then represented by a projection operator on ${\cal V}$. This construction
has the merit of preserving  the quantum logic structure \cite{I94} and highlighting
the  non-trivial temporal structure of histories \cite{Sav99a,Sav99b}. Furthermore,
one can also construct a Hilbert space ${\cal V}$ for continuous-time histories
\cite{IL95,ILSS98,An01b} by a suitable definition of the notion of the tensor product.

Furthermore, to each history $\alpha$ we may associate the class operator
$\hat{C}_{\alpha}$ defined by
\begin{equation}
\hat{C}_{\alpha} = \hat{U}^{\dagger}(t_n) \hat{\alpha}_{t_n} \hat{U}(t_n) \ldots
\hat{U}^{\dagger}(t_1) \hat{\alpha}_{t_1} \hat{U}(t_1).
\end{equation}

The decoherence functional is defined  as a complex-valued function of pairs of
histories: i.e. a map $d: {\cal V} \times {\cal V} \rightarrow {\bf C}$. For two
histories $\alpha$ and $\alpha'$ it is  given by
\begin{equation}
d(\alpha, \alpha') = Tr \left( \hat{C}_{\alpha} \hat{\rho}_0
\hat{C}_{\alpha'}^{\dagger} \right) \label{decfun}
\end{equation}
The consistent histories  interpretation of this object is that when $d(\alpha,
\alpha') = 0$ for $\alpha \neq \alpha'$ in an exhaustive and exclusive set of
histories \footnote{ By exhaustive we mean that at each moment of time $t_i$,
$\sum_{\hat{\alpha}_{t_i}} \hat{\alpha}_{t_i} = 1 $ and by exclusive that
$\hat{\alpha}_{t_i}  \hat{\beta}_{t_i} = \delta_{\alpha \beta}$. Note that by $\alpha$
we denote  the proposition with  the corresponding projector written as $\hat{\alpha}$
with a hat.}, then one may assign a probability distribution to this set as $p(
\alpha) = d(\alpha, \alpha)$. The value of $d(\alpha, \beta)$ is, therefore, a measure
of the degree of interference between the  histories $\alpha$ and $\beta$.

But one can view  the decoherence functional solely in an operational perspective.
The state $\hat{\rho}_0$ corresponds to a preparation of an ensemble of systems. This
can be visualised  as a {\em beam} of particles, which passes through various filters
in a succession. The filters correspond to the projection operators that constitute
the history; the diagonal elements of the decoherence functional gives the intensity
of the beam that has passed through these filters. There is no {\em a priori} reason
for these intensities to be described by an additive probability measure. The
decoherence condition can be taken as specifying the domain of validity of an
approximation, by which the full quantum mechanical description is substituted by an
effective one through probability theory \cite{An01c}. However, even the operational
description does not explain the physical meaning of the off-diagonal elements of the
decoherence functional.

\subsection{Interpretation of the off-diagonal elements }

The interpretation of the off-diagonal elements of the decoherence functional is more
conveniently carried out in terms of a version of the geometric phase, know as the
{\em Pancharatnam phase}.  Its origin lies in the following considerations.

 A quantum state is specified by a normalised Hilbert space vector, up to a phase.
 In other words,  a (pure) quantum state corresponds to an element of the projective
  Hilbert space $PH$. In fact, the unit sphere in the Hilbert space forms a $U(1)$
   principal fiber bundle over $PH$, known as the Hopf bundle.
   The absolute phase of a vector then makes
    no physical sense, but it is desirable to be able to compare the relative phase
     between two different  vectors.

  Consider two normalised vectors $|\psi \rangle$ and $| \phi
  \rangle$ and perform the operation of changing the phase of
  $| \psi \rangle$  by a factor $e^{i \chi}$. Interfering $e^{i \chi} |\psi
  \rangle$ and $ |\phi \rangle$, we get a beam with intensity
  \begin{equation}
|| e^{i \chi} | \psi \rangle + | \phi \rangle ||^2 = 2 + 2 | \langle \psi| \phi
\rangle | \cos (\chi -\arg \langle \psi| \phi \rangle)
\end{equation}
This intensity achieves a maximum for $\chi = \arg \langle \psi| \phi \rangle$. This
value for $\chi$  is the Pancharatnam phase between the two vectors $|\psi \rangle$
and $|\phi \rangle$. It has been experimentally determined in certain  occasions (see
\cite{WRFI97} for a measurement using neutron interferometry): one needs to implement
the transformations $| \psi \rangle \rightarrow e^{i \chi} | \psi \rangle$ (e.g
through the action of the Hamiltonian of
  which it $|\psi \rangle$ is an eigenstate) and then find the
  maximum of the intensity of the combined beam. We should note
  that the Pancharatnam phase has a close mathematical
  relationship with the natural connection on the Hilbert's space
  Hopf bundle \cite{SaBha88}.

  Consider, now, the simplest example of a history. We have a beam
  of particles characterised by the Hilbert space vector $| \psi
  \rangle $ and at times $t_1$ and $t_2$ measurements
  corresponding to (generally non-commuting operators) $\hat{\alpha}_1$ and $\hat{\alpha}_2$
  respectively. Let us absorb the Hamiltonian evolution in a
  redefinition of the projection operators. The beam passing
  through the projectors will be $\hat{\alpha}_2 \hat{\alpha}_1| \psi \rangle$. It is clear
   that the value of the decoherence functional between the
  history  $\alpha$ (that the system passed through $\alpha_1$ and
   $\alpha_2$ successively) and the trivial history $1$ (i.e. no measurement on $ | \psi
\rangle$)
  will be  the Pancharatnam phase between $|\psi
  \rangle$ and $ \hat{\alpha}_2 \hat{\alpha}_1|\psi \rangle$.

  Let us  be more precise. The number $\langle \psi| \hat{\alpha}_1 \hat{\alpha}_2 \hat{\alpha}_2 \hat{\alpha}_1| \psi
  \rangle$ can be determined by a measurement of the intensity of
  $\hat{\alpha}_2 \hat{\alpha}_1 |\psi \rangle$. It is essentially equal to the diagonal
  element of the decoherence functional $d(\alpha, \alpha) = r$.
  We want to measure $d(\alpha, 1)= \langle \psi| \hat{\alpha}_s \hat{\alpha}_1| \psi \rangle  := \rho
  e^{i \beta}$, where  $0 \leq \rho \leq 1$.

Similarly to the previous discussion we interfere
  the beam $\hat{\alpha}_2 \hat{\alpha}_1|\psi \rangle$, with $e^{i \chi} | \psi \rangle$.
  This yields the intensity
  \begin{equation}
I(\chi)  = || e^{i \chi} | \psi \rangle + \hat{\alpha}_2 \hat{\alpha}_1 |\psi \rangle
||^2 = 1 + r^2 +2 \rho \cos(\chi - \beta)
  \end{equation}

This intensity has a maximum for $\chi = \beta$: this determines $\arg d(\alpha,1)$.
At the maximum of the intensity we have $I_{max} = 1 + r^2 + 2 \rho$. Since we know
$r$ from a previous experiment and the value of $I_{max}$ can be measured, we can
determine $\rho$. Thus by measuring the intensity for different values of $\chi$, we
have determined the off-diagonal element of the decoherence functional. Note, that the
determination of this phase necessitates the study of the interference of two beams,
as is always the case in the measurement of quantum phases.

This procedure can be easily repeated -in principle- for the interference of arbitrary
pairs of histories. The off-diagonal elements of the decoherence functional will then
always be determined through the measurement of the Pancharatnam phase between two
vectors.

\subsection{Correlation functions}

Let us now consider an ensemble of quantum systems prepared in a state described by a
density matrix $\hat{\rho}$ and try to operationally construct the correlation
function of two observables $\hat{A} = \sum a_i \hat{\alpha}_i$ and $\hat{B} = \sum_j
b_j \hat{\beta}_j$
 at times $t_1$ and $t_2 > t_1$ respectively.
Here $\hat{\alpha}_i$ are an exhaustive  and exclusive  set of projectors,
 and so is $\hat{\beta}_j$.

Let the Hamiltonian of the system be $\hat{H}$ and $\hat{\rho}_0$ the state of the
system at time $t = 0$. The probability that both $\hat{\alpha}_i$ {\em and then }
$\hat{\beta}_j$ are true  will be
\begin{eqnarray}
p(i, t_1; j, t_2) = Tr \left( \hat{\beta}_j e^{-i\hat{H}(t_2 - t_1)}\hat{\alpha}_i
 e^{-i\hat{H}t_1 } \hat{\rho}_0 e^{i\hat{H}t_1 } \hat{\alpha}_i e^{i
\hat{H}(t_2 - t_1)} \right) =
\nonumber \\
Tr \left( \hat{\beta}_j(t_2)\hat{\alpha}_i(t_1)\hat{\rho}_0 \hat{\alpha}_i(t_1)
\right),
\end{eqnarray}

 If we now  vary over
all possible values of $i$ and $j$, we can construct the {\em statistical} correlation
function between $\hat{A}$ and $\hat{B}$
\begin{equation}
\langle \hat{A}_{t_1} \hat{B}_{t_2} \rangle_S = \sum_{ij} a_i b_j p(i. t_1; j, t_2)
\label{statcor}
\end{equation}

But this correlation function is {\em not} what one usually calls correlation function
in quantum theory. This name is usually employed for the expectation of a product of
operators
\begin{equation}
\langle \hat{A}_{t_1}\hat{ B}_{t_2} \rangle = Tr \left( \hat{A}(t_1) \hat{B}(t_2)
\hat{\rho} \right) = \sum_{ij} a_i b_j Tr \left(
 \hat{\alpha}_i(t_1)  \hat{\beta}_j(t_2) \hat{\rho} \right).
\label{qcor}
\end{equation}
This  is a complex-valued object, in contrast to (1.10) that was constructed using
 frequencies of events and can only
be real-valued. What does then the quantum mechanical correlation correspond to? Our
previous discussion makes it now clear. The complex-valued quantum mechanical
correlation function is related to the off-diagonal elements of the decoherence
functional and essentially contains information about relative phases.

The precise relation is as follows.
 Let $\hat{A}^a$
denote a family of commuting operators. Then the time-ordered two-point correlation
function is defined as
\begin{eqnarray}
G^{2,0}(a_1,t_1;a_2,t_2) = \theta(t_2-t_1) Tr[
 \hat{A}^{a_1}(t_1) \hat{A}^{a_2}(t_2) \hat{\rho}_0] + \nonumber \\
 \theta(t_1-t_2) Tr[ \hat{A}^{a_2}(t_2) \hat{A}^{a_1}(t_1)  \hat{\rho}_0]
\end{eqnarray}
One can similarly define time-ordered $n$-point functions, or anti-time-ordered
\begin{eqnarray}
G^{0,2}(a_1,t_1;a_2,t_2) = \theta(t_1-t_2)
 Tr[ \hat{A}^{a_1}(t_1) \hat{A}^{a_2}(t_2)  \hat{\rho}_0] + \nonumber \\
 \theta(t_2-t_1) Tr[  \hat{A}^{^a_2}(t_2) \hat{A}^{a_1}(t_1)  \hat{\rho}_0]
\end{eqnarray}
In general, one can define {\em mixed} correlation functions $G^{r,s}$, with $r$
time-ordered and $s$ anti-time-ordered entries, as for instance
\begin{eqnarray}
G^{2,1} (a_1,t_1; a_2,t_2|b_1, t'_1) = \theta(t_2-t_1) Tr[ \hat{A}^{a_1}(t_1)
\hat{A}^{a_2}(t_2)  \hat{\rho}_0 [\hat{A}^{b_1}(t_1') ] +
\nonumber \\
 \theta(t_1-t_2) Tr[ \hat{A}^{a_2}(t_2) \hat{A}^{a_1}(t_1) \hat{\rho}_0
\hat{A}^{b_1}(t_1')]
\end{eqnarray}
These correlation functions are generated by the Closed-Time-Path (CTP) generating
functional associated to the family $\hat{A}^a$
\begin{eqnarray}
Z_A[J_+,J_-] = \sum_{n,m=0}^{\infty} \frac{{i}^n (-i)^m}{n! m!} \int dt_1 \ldots dt_n dt'_1 \ldots dt'_m  \nonumber \\
G^{n,m}(a_1,t_1; \ldots a_n,t_n | b_1,t'_1;\ldots; b_m,t'_m) \nonumber \\
J_+^{a_1}(t_1) \ldots J^{a_n}(t_n) J_-^{b_1}(t'_1) \ldots J^{b_m}(t'_m)
\end{eqnarray}
The name closed-time arose, because in the original conception (by Schwinger
\cite{Schw61}
 and Keldysh \cite{Kel64} the time path one follows is from some initial time
$t=0$ to $t \rightarrow \infty $ moving in a  time-ordered fashion and then back from
infinity to $0$ in an anti-time-ordered fashion. The total time-path is in effect
closed.

Clearly there must be  a relation between the decoherence functional and the CTP one.
One can see in the correlation functions, if we assume a single operator  $\hat{A} =
\sum_i a_i \hat{\alpha}_i$ and consider a pair of histories $\alpha(i_1,t_1; \ldots;
i_n,t_n) = \{\hat{\alpha}_{i_1},t_1; \ldots; \hat{\alpha}_{i_n},t_n \} $ and
$\alpha'(i_1,t'_1; \ldots;i_n,t'_m) = \{\hat{\alpha}_{j_1},t'_1; \ldots;
\hat{\alpha}_{j_m},t_m \}$. Then one can  verify \cite{An01b}that
\begin{eqnarray}
G_A^{n,m}(t_1,\ldots, t_n;t'_1,\ldots, t'_m) = \sum_{i_1\ldots i_n} \sum_{j_1 \ldots
j_m} a_{i_1}\ldots a_{i_n}
 b_{j_1} \ldots b_{j_m} \nonumber \\
\times d[\alpha (i_1,t_1; \ldots; i_n,t_n),\alpha' (j_1,t_1; \ldots; j_m,t_m)]
\end{eqnarray}

 One  needs to consider a decoherence
functional for continuous-time histories \cite{IL95,ILSS98, Sav99a}  and this requires
a significant upgrading of the formalism of quantum mechanical histories. The key idea
 is to represent histories by projectors on a tensor product of Hilbert spaces
 $\otimes_{t \in T} H_t$ \cite{I94} in analogy to the
construction of the history sample space classically. A suitable Hilbert
 space (not  a genuine tensor product) can be constructed \cite{IL95}
for the case that $T$ is a continuous set and the decoherence functional can be
defined as a bilinear, hermitian functional on this space. It can then be shown that
as a functional it is essentially a double
 "Fourier transform" of the CTP generating functional \cite{An01b}. This is a
 construction, we shall repeat in the context of the present
 paper.

\subsection{Commutative observables}

The study of quantum mechanical histories has shown that phases are equally  important
structural elements of quantum theory to probabilities and they are closely related to
the non-additivity of the probability assignment of quantum theory. Moreover, there
exists  a well defined operational procedure by which they can be measured. Accepting
phases as primitive elements of quantum theory allows us to phrase it in a way that is
independent of Kolmogorov' probability or their generalisations.

Now the assumption of Kolmogorov probability is a crucial ingredient of
(generalisations of) proofs of Bell's theorem against the possibility of local hidden
variable theories. An assumption that follows from additive probability (namely that
if an event has probability one its complement has probability zero) is employed in
the Kochen-Specker's theorem against uncontextual assignment of properties in quantum
systems. If the assumption of Kolmogorov probability is abandoned the way is open to
write versions of quantum theory that contain local hidden-variable theories.

Such theories are, of course, non-deterministic and not probability theories in the
classical sense. The behaviour of individual systems is ``random'' and ensembles
exhibit behaviour that manifests in non-additive probabilities and relative phases.
This information can be included in a version of the decoherence functional.

In effect, such are history theories (satisfying a modified  form of the
Gell-Mann-Hartle-Isham axioms), which are described by {\em commutative} observables.
The basic observables are then functions on some classical set, usually a manifold. In
\cite{An01a} we argued that the best choice for this base space would be the classical
phase space. The reason is that  the quantum mechanical Hilbert space typically
contains the irreducible representation of a group, which also acts transitively on
the classical phase space (see \cite{I83} for the general perspective on this
quantum-symplectic correspondence). This provides  a way (in fact more than one) to
map quantum mechanical operators to functions on the phase space. A theory developed
on phase space would then be able to include all information included in the Hilbert
space operators. This we showed to be true using the Wigner transform.

The main difference of such theories from standard quantum mechanics lies in the
notion of {\em an  event}. In a classical space a (sharp) event corresponds to a
subset of the phase space, while in quantum theory events correspond to projection
operators. There exists no exact correspondence between these two objects; however the
two theories give identical quantum mechanical correlation functions of the type
(\ref{qcor}) and their difference is at the order of $\hbar$ in the statistical
correlation functions (\ref{statcor}). We have not been able to find any experimental
situation that would be able to unambiguously distinguish them.

In the following sections we shall give a more detailed construction of  this class of
theories.

\section{The basic structures}

We shall here provide the basic features of the class of theories we described
earlier; they are based on the introduction of a decoherence functional for paths on
the classical phase space. We shall draw analogies on the formal similarity with
stochastic processes; for this reason we shall refer to  our class of theories as
theories of quantum processes.

\subsection{Events and observables}
At the level of observables, the structure of our theory is identical with that  of
classical probability theory. That is, we assume the existence of a space $\Omega$ of
elementary alternatives. A point of $\Omega$ corresponds to the most precise
information one can extract from a measurement of the quantum system. Note, that at
this level we do not distinguish whether $\Omega$ refers to properties of a systems at
one moment of time or to histories. Our definitions are general and only in the next
section shall we specify the history content.

 This space $\Omega$ has to be equipped with some
additional structure. In general, a measurement will yield some information that the
system was found in a given subset of $\Omega$. But not all subsets of $\Omega$ are
suitable to incorporate measurement outcomes.
 For instance, when we consider position it is physically meaningless to
 consider the subset of rational values of position (with respect to some unit).
 One, therefore needs to choose a family of subsets ${\cal C}$ of $\Omega$, that correspond to
 the coarse-grained information we can obtain about the physical systems. These sets are often
  called {\em events}. The   family ${\cal C}$ containing the events  has to satisfy some
   natural mathematical conditions: in mathematical terms ${\cal C}$ has to be a $\sigma$-field.
    The relevant conditions  are the following \\ \\ \\
 A1. $\Omega \in {\cal C}$: if an experiment is performed one of the outcomes will occur. \\ \\
A2. $\emptyset \in {\cal C}$: it is impossible that no outcome results if an
experiment
 is performed. \\ \\
A3. If $A \in {\cal C}$, then $\Omega - A \in {\cal C}$: if $A$ is a possible
measurement
 outcome then so can be its complement. \\ \\
A4. If $A, B \in {\cal C}$, then $A \cup B \in {\cal C}$ and $A \cap B \in {\cal C}$: unions and intersections of experimental outcomes are also possible  experimental outcomes. \\ \\
A5. For countably many $A_n \in {\cal C}, n = 1, 2, \ldots$, $\cup_{n=1}^{\infty} \in
{\cal C}$. This is a technical condition particularly relevant when dealing with
 the case where $\Omega$ is a manifold.
  \\ \\
Equipping $\Omega$ with a $\sigma$-field turns it into a {\em measurable space}. We
shall often focus on a particular set of events $A_i$.  By repeatedly applying the
operations of countable union, intersection and complementation we can construct a
$\sigma$-field {\em generated by the collection $A_i$}. This $\sigma$-field will be
denoted as $\sigma(A_i)$.
\\ \\
{\bf Some examples} \\
1. If we take a single set $A$, then $\sigma(A) = \{ \emptyset, \Omega, A, \Omega -A
\}$.
 \\ \\
2. One can also consider the $\sigma$-field that is generated by a set of mutually
disjoint subsets $A_i$ of $\Omega$, that are also exclusive. This essentially
represents  to a pointer device, each $i$ a position of the pointer. This field will
be denoted as $\sigma(\{A\})$.
\\ \\
3. If $\Omega$ is a topological space, we usually consider the $\sigma$-field
generated by all {\em closed} subsets of $\Omega$. This is known as the Borel
$\sigma$-field  ${\cal B}(\Omega)$. If we consider the real line ${\bf R}$, clearly
the corresponding Borel field  ${\cal B}({\bf R})$ (or ${\cal B}$ for short) is
generated by all interval  $[a,b], a, b \in {\bf R}$. We shall mostly assume our
$\sigma$-fields to be Borel fields.
\\ \\
4. If we have a function $F: \Omega \rightarrow {\bf R}$, we can define the
$\sigma$-field  generated by $F$ as the one generated of all sets of the form
$F^{-1}(B)$, where $B$ is  a Borel set on ${\bf R}$.
\\
\\
An observable is what is actually determined in an experiment. Since in experiments we
eventually come to measure real numbers (or occasionally integers, which can be
embedded into the real numbers) the mathematical object that would represent the
notion of observable is a map from $\Omega$ to ${\bf R}$. However, not all possible
maps will do:
 the structure of the $\sigma$-fields has to be preserved.
  This amounts to the following condition. \\ \\
If $f: \Omega \rightarrow {\bf R}$ and $C \in {\cal B}$, then $f^{-1}(C) \in {\cal
C}$.
\\ \\
Such functions are called {\em measurable} and in the language  of probability theory
are known as  {\em random variables}. We shall denote the space they belong to in as
$F(\Omega)$. In the following, unless it is explicitly specified, any function we will
refer to, will be an element of $F(\Omega)$.

Among all functions, important are characteristic functions of the various  subsets of
$\Omega$. These are defined as
\begin{eqnarray}
\chi_A(x) &=& 1 , x \in \Omega \\
 &=& 0 , x \notin \Omega
\end{eqnarray}

An important property of the characteristic functions is the following. If $\lambda$
is a possible value of a random variable $f$ and  $A_{\lambda} = f^{-1}({\lambda})$,
then it is evident
\begin{equation}
f = \int d \lambda \lambda \chi_{A_{\lambda}}.
\end{equation}

This  relationship is the prototype of the most important spectral theorem in the case
of Hilbert space observables and its analogue in the commutative case.

\subsection{The decoherence functional}

So far our construction has (deliberately) been identical to that of  classical
probability theory. Now we come to the point of departure. For classical probability
theory this is the point to define a probability measure. Let us for the sake of
completeness and
later comparisons  give its standard definition. \\ \\ \\
A {\em probability measure} $p$ on $\Omega$ is a map $p: {\cal A} \rightarrow [0,1]$,
such that
\\
- $p(\emptyset) = 0$, ({\em null triviality})  \\
- $p(\Omega) = 1$, ({\em normalisation})  \\
- for $A, B \in {\cal C}$, such that $A \cap B = \emptyset$, $p(A \cup B) = p(A) +
p(B)$, ({\em additivity}).

Note, that the additivity  condition is often strengthened to include  {\em countable}
unions of mutually disjoint sets $A$. Clearly this definition implies that  $p$ can be
extended to any measurable function  on $\Omega$
\\ \\
Now in quantum theory (and more particularly histories), one measures  relative phases
in addition to probabilities. As we explained earlier the object that incorporates
them is the decoherence functional. One then needs to give some basic properties or
axioms by which a decoherence functional will be defined. This has been done by
Gell-Mann and Hartle \cite{Har93a} and subsequently generalised by  Isham and Linden
\cite{I94, IL94}, in the context of consistent histories. Their construction lay
within the consistent histories approach. They, therefore, considered more general
structures for the field of events. Our scheme here is a adoption of theirs in our
present context.

Before we give the axioms for the decoherence functional, we should  remark that the
symbol $D$ or $d$ canonically used in the literature (and employed in the previous
section)  is very unwieldy and prone to confusion  with differentials. For this
reason, we took the liberty to denote the decoherence functional as $\Phi$ (for phase)
to denote its most important role as the carrier of the phase information.
\\ \\ \\
A {\em decoherence functional} $\Phi$ is a map from ${\cal C} \times {\cal C}
\rightarrow  {\bf C}$, such that the following conditions are satisfied
\\ \\ \\
B1. {\em Null triviality:} For any $A \in {\cal C}$, $\Phi(\emptyset,A) = 0$. \\
 In terms of our interpretation of the off-diagonal elements of the decoherence functional
 as corresponding to Pancharatnam phases, there can be no phase measurement if one
 of the two beams that have to be interfered is absent. \\ \\
\\
B2. {\em Hermiticity:} For $A,B \in {\cal C}$, $\Phi(B,A) = \Phi^*(A,B)$. \\
Clearly the phase difference between two histories becomes opposite if we exchange the
 sequence, by which these histories are considered.
\\ \\ \\
B3. {\em Positivity:} For any $A \in {\cal C}$, $\Phi(A,A) \geq 0$. \\
This amounts to the fact that the diagonal elements of the decoherence functional
 are interpreted as probabilities (albeit non-additive). Operationally probabilities
  are defined by the number of times a particular event occurred in the ensemble and  by
   definition they can only be positive.
\\ \\ \\
B4. {\em Normalisation:} $\Phi(\Omega, \Omega) = 1$. \\
Clearly, if no measurement takes place the intensity of the beam would never change.
\\ \\ \\
B5. {\em Additivity:} If $A, B, C \in {\cal C}$ and $A \cap B = \emptyset$,  then
$\Phi(A \cup B,C) =
\Phi(A,C) + \Phi(B,C)$. \\
There is no intuitive operational reason, why this should be the case. This property
is equivalent to the superposition principle of quantum theory and we can consider
that it is forced upon us by experimental results. Of course, this is the property
that makes the decoherence functional the natural object to use. Note, that it is
probably natural to extend it for countable unions of mutually disjoint subsets $A_i$.
\\ \\ \\
B6. {\em Boundedness:} For all $A,B \in {\cal C}$, $ |\Phi(A,B)| \leq 1 $. \\
This is suggested by the measurability of our general discussion in section 2.2In
standard quantum theory this condition is  satisfied by homogeneous histories
\footnote{The proof for a decoherence functional of the form (2.2) is straightforward:
\begin{equation}|\Phi(\alpha, \beta)| = |Tr \left(\hat{C}_{\alpha} \hat{\rho}_0 \hat{C}_{\beta}^{\dagger} \right)| \leq Tr|(\hat{C}_{\alpha} \hat{\rho}_0 \hat{C}_{\beta}^{\dagger}| \leq Tr \hat{\rho}_0 ||\hat{C}_{\alpha}|| || \hat{C}_{\beta}^{\dagger}|| \leq 1 . \nonumber
\end{equation}},
i.e histories that can be written in the form $(\alpha_1, t_1; \alpha_2, t_2; \ldots ;
\alpha_n, t_n)$. It is not true  for their generic conjunctions, when viewed  as
propositions. These  conjunctions  are called {\em inhomogeneous histories} in
\cite{I94,IL94}. When the lattice of propositions is Boolean like in our case all
conjunctions of homogeneous histories are homogeneous histories and we can expect this
condition to be valid.
\\ \\ \\
Now a decoherence functional $\Phi$ can be extended into a functional over
$F(\Omega)$.  One can clearly define its values on a pair of functions $F,G \in
F(\Omega)$ by additivity
\begin{equation}
\Phi(F,G) :=  \int d \lambda d \mu \lambda \mu \Phi(F^{-1}(\{\lambda\}), G^{-1}(
\{\mu\})),
\end{equation}
where $\lambda$ and $\mu$ are  possible real values for the functions $f$ and $g$
respectively.

We keep the same symbol $\Phi$ for the decoherence functional, viewed as functional
either on ${\cal C}$ or on $F(\Omega)$. The properties 1-6 are then translated as
follows:
\\ \\
C1. {\em Null triviality:} For any $F \in F(\Omega)$, $\Phi(0,F) = 0$. \\
C2. {\em Hermiticity:} For any $F,G \in F(\Omega)$, $\Phi(G,F) = \Phi^*(F,G)$. \\
C3. {\em Positivity:} For any $F \in F(\Omega)$ , $\Phi(F,F) \geq 0$ \footnote{ This
property is satisfied in quantum theory, since by additivity we have $\Phi(F,F) =
\sum_{\alpha, \beta} \lambda_{\alpha} \lambda_{\beta} \Phi(A_\alpha, A_\beta)$, where
we wrote the spectral decomposition $F = \sum \lambda_{\alpha} \chi_{A_{\alpha}}$.
Using equation (2.2) let us write $\hat{A} = \sum_{\alpha} \lambda_{\alpha}
\hat{C}_{\alpha}$, then $\Phi(F,F) = Tr \left(\hat{\rho}_0 \hat{A}^{\dagger} \hat{A}
\right) \geq 0$. In fact, for complex valued $F$ we have by the same token
$\Phi(\bar{F},F) \geq 0$.}
. \\
C4. {\em Normalisation:} $\Phi(1,1) = 1$.\\
C5. {\em Additivity:} For any $F,G,H \in F(\Omega)$, $\Phi(F,G+H) = \Phi(F,G) + \Phi(F,H)$. \\
C6. {\em  Boundedness:} For any $F,G \in F(\Omega)$, $|\Phi(F,G)| \leq ||F|| ||G||$,
where $||\cdot ||$ denotes the supremum norm on $\Omega$.
\\ \\
Whenever $\Omega$ is a space  equipped with a measure $dx$ (of the Lebesque type), we
can focus our attention to decoherence functionals   that can be
 obtained by a density: i.e.  we can write
\begin{equation}
\Phi(F,G) = \int  F(x) G(x') \Phi(dx,dx') = \int dx dx' F(x) G(x') \upsilon(x,x')
\end{equation}
in terms of a complex-valued density $\upsilon$ on $\Omega \times \Omega$. In a
precise mathematical language  $\upsilon(x,x')$ is the Radon-Nikodym derivative of
$\Phi(\cdot,\cdot)$ with respect to the Lebesque measure $dx \otimes dx$ on $\Omega
\times \Omega$. This amounts to viewing   $\Phi$
 as a complex-valued measure on $\Omega \times \Omega$. This perspective will be very useful in the
mathematical development of our theory as we can exploit  a number  formal
similarities with the complex measures.

In particular, we can employ the powerful Radon-Nikodym theorem, which states that if
$\nu$ is a $\sigma$-additive measure over a $\sigma$-field over which another
$\sigma$-additive measure $\mu$ is define, then up to a set of measure zero we can
write a random variable  $f$ such that $\nu(A) = \int_A f d \mu$. This function $f$ is
the Radon-Nikodym derivative, which was employed earlier to define the density
$\upsilon$. This theorem forms the basis of the formulation of conditional
expectations.

We should also remind the reader that the axioms 1-6 for the decoherence functional
are essentially the properties that a density matrix would have if it was viewed as a
bilinear kernel over a space $\Omega$. It is in this sense that the decoherence
functional is a generalisation of the density matrix in the history setting.

A pair of events for which $\Phi(A,B) = 0$ is called a pair of measure zero, while
any event $A$ for which $\Phi(A,B) = 0$, for all $B$ is called an {\em event of
measure zero}.

\subsection{Simple examples}
Let us now give some examples by which  a decoherence functional appears in quantum
theory.  Our focus is eventually histories; however, at this point we shall consider
systems at a single moment of time, mostly in order to make the connection with the
standard formalism of quantum theory more clear. We then seek to see how a decoherence
functional of the form $Tr (\hat{A} \hat{\rho} \hat{B})$ can be represented in terms
of commutative observables
\\ \\ \\
{\bf Example 1:} Let us consider the case that $\Omega = {\bf R}$ corresponds  to the
spectrum of a self-adjoint operator $A$. If $\rho$ is the density matrix of the system
then the distribution function for the decoherence functional is
\begin{equation}
\upsilon(x,x') = \rho(x,x') \delta(x,x')
\end{equation}
Then for any Borel subsets of ${\bf R}$ $A, B$ we get the trivial result (no complex
phases)
\begin{equation}
\Phi(A,B) = \int_{A \cap B} dx \rho(x,x) = p(A \cap B)
\end{equation}
\\ \\ \\
{\bf Example 2:} More interesting is the case where $\Omega$ is taken to be the
classical  phase space and the correspondence between quantum operators and classical
objects is taken through the Wigner transform. Let us consider for simplicity an one
dimensional system, characterised by the canonical commutation relation
\begin{equation}
[q,p] = i
\end{equation}
We can write the Weyl operator $\hat{U}(x,\xi) = e^{i x \hat{p} + i \xi \hat{q}}$.
The Wigner transform provides a map from any function on the system's Hilbert space
and the classical phase space:
\begin{equation}
\hat{A} \rightarrow F_{\hat{A}} = \int dx d\xi e^{-iq \xi - i p x} Tr(\hat{A}
\hat{U}(x, \xi)): = Tr(\hat{A} \hat{\Delta}(q,p),
\end{equation}
The density corresponding  $\rho$ is then
\begin{equation}
\upsilon(q,p;q'p') = Tr \left( \hat{\Delta}(q,p) \rho \hat{\Delta}(q',p') \right)
\end{equation}
Using the fact that $\hat{U}(x,\xi)  \hat{U}(x',\xi') = e^{i (\xi x' - \xi' x)}
U(x+x',\xi + \xi')$ we derive that
\begin{equation}
\upsilon(q,p;q'p') = e^{i( qp'-pq')} W(q'-q,p-p'),
\end{equation}
where $W(q,p) = Tr  \left( \hat{\rho} \hat{\Delta}(q,p) \right)$ is the Wigner
function of the classical system. So if we consider two subsets of the phase space $A$
and $B$ we have
\begin{equation}
\Phi(A,B) =  \int_A dq dp \int_B dq' dp' e^{i( qp'-pq')} W(q'-q,p-p')
\end{equation}
It is clear that the probability $\Phi(A,A)$ is not given by the Wigner  function as a
probability distribution. This corresponds to the well known fact that the Wigner
function is not a genuine probability distribution, since it is negative. We can get a
classical probabilistic description only if we consider that the volume of $A$ and $B$
is much larger than unity (remember $\hbar = 1$), in which case some smeared
(positive) version of the Wigner function becomes a probability distribution. It is
worth remarking that if we are able to produce devices that correspond to accurate
phase space measurements, then one would be able to determine an interference phase
even in a single moment of time.
\\ \\ \\
{\bf Example 3:} An alternative description of phase space properties is  by means of
the coherent states. If we define the coherent states as
\begin{equation}
| x, \xi \rangle = \hat{U}(x,\xi) |0 \rangle,
\end{equation}
where $|0 \rangle$ is a reference state, often taken to be the ground state of the
system's  Hamiltonian. Using the coherent states to describe the classical phase space
we have
\begin{equation}
\upsilon(x,\xi;x',\xi') = \langle x \xi|\rho| x' \xi' \rangle \langle x' \xi' |x \xi
\rangle
\end{equation}

\subsection{Combination of subsystems}
If we have two separate systems, characterised by sample spaces $\Omega_1$ and
$\Omega_2$, then the combined system is characterised by the Cartesian product
$\Omega_1 \times \Omega_2$. If also the two systems are independent and characterised
by decoherence functionals $\Phi_1$ and $\Phi_2$ respectively, then the combined
system is described by the decoherence functional $\Phi_1 \otimes \Phi_2$, which is
defined as
\begin{equation}
(\Phi_1 \otimes \Phi_2)(A_1 \times A_2, B_1 \times B_2) = \Phi_1(A_1,B_1) \times
\Phi_2(A_2, B_2).
\end{equation}

\subsection{Conditioning}
\subsubsection{The classical case}
Conditioning it is a very important part of classical probability;
 it is the mathematical implementation of the idea that when we obtain
 information from an experiment, we need to modify the way we describe the system
 (i.e. the probability distribution) in order to account for the new information.
 The prototype of conditioning classically is the notion of conditional
 probability, i.e. the probability that $A$ will take place when we have
 verified that $B$ occured. Writing the conditional probability as
  $p(A|B)$ we have that
\begin{equation}
p(A|B) = \frac{p(A \cap B)}{p(B)} \label{condprop}
\end{equation}
Now this is a very restrictive definition of conditional probability, since it cannot
be  generalised to the case that $p(B) = 0$; objects like propagators are defined
through conditional probabilities of events with zero measure.

The most general possible definition that allows such generalisations  is that of
conditional expectation. This refers to a $\sigma$-field, which is a subfield ${\cal
A}$ of ${\cal C}$. The elements of the subfield ${\cal A}$ correspond to the events we
want to condition.

In classical probability the conditional expectation of a random variable  $F \in
F(\Omega)$  with respect to ${\cal A}$ is defined as a {\em random variable} denoted
as $p(F|{\cal A})$ that is measurable with respect to ${\cal C}$ and is such that for
all $B \in {\cal A}$
\begin{equation}
\int_{B} p(dx) p(F|{\cal A}) = \int_{B} F(x) p(dx)
\end{equation}
The existence of such variables can be proved by virtue of the Radon-Nikodym  theorem
for any subfield ${\cal C}$. It is unique up to a set of measure zero. To understand
the nature of this object $p(F|{\cal A})$ consider the case of the subfield
corresponding to a pointer device $\sigma (\{A \})$. If $p(A_i) \neq 0$ then we have
\begin{equation}
p(F|\sigma( \{A \}))(x) = \sum_i \frac{p(F \chi_{A_i})}{p(A_i)} \chi_{A_i}(x)
\label{condpointer}
\end{equation}
However, the definition of $p(F| \sigma(\{ A \} )$ makes sense also for cases  where
$p(A_i) = 0$ for some values of $i$. One just drops these values from the summation in
(\ref{condpointer}).

If $F = \chi(C)$ for some subset $C$ of $\Omega$ we define the conditional probability
as a random variable $p(C|{\cal C}) = p(\chi_C|{\cal A})$ and in the case of equation
(\ref{condpointer})
\begin{equation}
p(C|\sigma(\{ A \}) )(x) = \sum_i \frac{p(C \cap A_i)}{p(A_i)} \chi_{A_i}(x)
\end{equation}
Note that for a fixed value of $x \in A_i$, which refers to a definite experimental
outcome we have
\begin{equation}
p(C|\sigma(\{ A \}) =  \frac{p(C \cap A_i)}{p(A_i)},
\end{equation}
which amounts to equation (\ref{condprop}) for conditional probabilities. However, it
is not always true  that we can define a probability distribution out of the fixed
choice of the variable $x$.

\subsubsection{The quantum case}
The above-mentioned logic can be straightforwardly generalised for conditioning with
respect to the decoherence functional, since this can be viewed as a complex measure
on $\Omega \times \Omega$. The different physical meaning of its contents, though, we
lead us to distinctions that are absent in the classical case.

First, given a subalgebra ${\cal A}$ of ${\cal C}$, we can define an object that is
mathematically analogous to the conditional expectation. That is, for a pair of random
variables $F$ and $G$ we can define a function on $F_C(\Omega) \otimes F_C(\Omega)$,
which we shall write us  $\tilde{\Phi}(F,G|{\cal C})$ and call the {\em conditional
pair}. This  is measurable on ${\cal A} \times {\cal A}$ and for all $B, B' \in {\cal
A}$
\begin{equation}
\int_{B \times B'} \tilde{\Phi}(F,G|{\cal A})(x,x')  \Phi(dx,dx') = \int_{B \times B'}
F(x) g(x') \Phi(dx,dx') \label{condpair}
\end{equation}
Unlike the classical case $\tilde{\Phi}(F,G|{\cal A})$ is not a random variable, not
even  a pair of random variables as it is correlated with respect to the two copies of
$\Omega$ entering its arguments. For the case of the pointer sub-field  and when
$\Phi(A_i,A_j) \neq 0 $ for all $A_i, A_j$ it reads
\begin{equation}
\tilde{\Phi}(F,G|\sigma(\{A \}))(x,x')  = \sum_{ij} \frac{\Phi(F \chi_{A_i}, G
\chi_{A_j})}{\Phi(A_i,A_j)} \chi_{A_i}(x) \chi_{A_j}(x') \label{condpairpointer}
\end{equation}
An important property of the conditional pair is the following:
\\ \\
If ${\cal A'} \subseteq {\cal A}$, then we have that
\begin{eqnarray}
\tilde{\Phi}[\tilde{\Phi}(F,G|{\cal A})|{\cal A'}] =
\tilde{\Phi}[\tilde{\Phi}(F,G|{\cal A'})|{\cal A}] = \tilde{\Phi}(F,G|{\cal A'}),
\label{condtwo}
\end{eqnarray}

 This means that the conditioning with respect to the more
stringent field ${\cal A'}$ prevails. The proof follows from the definition of
$\tilde{\Phi}$ and can be intuitively viewed by examining the case of pointer
subfields.

Another important property is that normalisation is preserved by conditioning, in the
sense that
\begin{equation}
\tilde{\Phi}(1,1|{\cal A}) = 1 \label{condnorm}
\end{equation}

The conditional pair is the correct mathematical way to implement the notion of
conditioning for decoherence functionals. It naturally  arises from the similarity of
$\Phi$ to a complex-valued measure. It   has, however,  little direct physical meaning
as it stands. For instance, taking a fixed value for the arguments $x$ and $x'$ does
not give a decoherence functional (unlike the classical case), because it does not
satisfy the positivity condition.

However, we can concentrate on the diagonal elements of the conditional pair by
defining
\begin{equation}
\Phi(F,G|{\cal C}) = \int dx' \delta(x,x') \tilde{\Phi}(F,G|{\cal A})(x,x')
\label{condcorr}
\end{equation}

$\Phi(F,G|{\cal C})$ {\em is} a complex valued random variable, which we shall call
the {\em conditional correlation} of $F$ and $G$. In the case of the pointer subfield
we have
\begin{equation}
\Phi(F,G|\sigma(\{A\}))(x) = \sum_i \frac{ \Phi(F \chi_{A_i},G
\chi_{A_i})}{\Phi(A_i,A_i)} \chi_{A_i}(x) \label{condcorrpointer}
\end{equation}
For a fixed value of $x \in A_i$ (which amounts to a determination of what the pointer
actually showed)  this gives a conditioned decoherence functional
\begin{equation}
\frac{\Phi(F \chi_{A_i}, G \chi_{A_i})}{\Phi(A_i,A_i)},
\end{equation}

If the $A_i$ corresponds to a subset of the spectrum of some operator it will be
represented in quantum theory by a projection operator $\hat{P}$ (as in example 1)
this gives the familiar {\em reduction of the wave packet} rule
\begin{equation}
\hat{\rho} \rightarrow \frac{\hat{P} \hat{\rho} \hat{P}}{Tr ( \hat{\rho} \hat{P})}.
\end{equation}

For the case  that the $\sigma$-field is generated by a family of functions $Z^a$, $a
= 1, \ldots n$, the corresponding conditioned pair will be denoted as
$\tilde{\Phi}(F,G|Z^a)$. This is a measurable function on $\Omega \times \Omega$. By
construction it is measurable with respect to all sets of the form $(Z^a)^{-1}(B)
\times (Z^b)^{-1}(B')$, for $B$, $B'$ Borel sets in ${\bf R}$. It can therefore be
written as $h(Z,Z')$, where $h: {\bf R}^n \times {\bf R}^n \rightarrow {\bf C}$ is a
measurable map. We can then define the quantity
\begin{equation}
\tilde{\Phi}(F,G|Z^a = z^a, Z^b = z'^b): = \tilde{\Phi}(F,G|z^a,z'^b) : = h(z,z').
\end{equation}
This object will be of great importance in the discussion of the Markov property.

Note that by the same token one can construct a conditional correlation
$\Phi(F,G|z^a)$.

\section{Quantum processes}

\subsection{Basic structures}
\subsubsection{Space of histories}
Now we shall consider the case, where the space $\Omega$ of events corresponds to
histories. We take $\Omega$ to be a space of paths from the set $T$, where the time
parameter lies, to a manifold which corresponds to the fine grained alternatives at a
single moment of time. The set $T$ will typically be either a closed interval of the
real line or the whole real line. The space $\Gamma$ will be taken as the classical
phase space of the system and its points will be denoted usually as $z$. The space of
histories $\Omega$ will be taken to consist of maps from $T$ to $\Gamma$, that are at
least continuous. Note that $\Omega$  is a subset of $\times_t \Gamma_t$, where
$\Gamma_t$ is a copy of $\Gamma$ defined at time $t$. The histories -elements of
$\Omega$- will be denoted as $z(\cdot)$.

The $\sigma$-field on $\Omega$ will be generated by the  Borel subsets of $\Gamma_t$
for all times $t$. Now if $f$ is a measurable function on $\Gamma$, one can define the
one-parameter family of measurable functions $F_t$ on $\Omega$ as
\begin{equation}
F_t(z(\cdot)) = f(z(t)) \label{canhis}
\end{equation}

A {\em quantum process} is defined as a triple $(\Omega, \Phi, F^a_t)$, where $\Omega$
is a history space as previously defined, $\Phi$ is a decoherence functional and
$F^a_t$ is family of functions indexed by $t$. Usually for $F^a_t$ we will consider
functions of the form (\ref{canhis}), with corresponding $f^a$ a set of coordinates on
$\Gamma$, or the generators of a group acting transitively on $\Gamma$ (as $\Gamma $
is taken to be a symplectic manifold).

\subsubsection{The decoherence functional}
The first issue that concerns us, is how a decoherence functional can be defined on
the path space $\Omega$. In \cite{An01b} we briefly explained  that this can be
achieved  in a similar fashion as probability measures on path spaces are defined in
the theory of stochastic processes. This procedure is as follows:

Let us consider a subset $\Delta_n$ of $T = [0, \tau]$ consisting of $n$-time points
$\{t_1, t_2, \ldots, t_n \}$ and define the space $\Gamma^{\Delta_n} = \times_{t \in
\Delta_n} \Gamma_t$. The map $i_{\Delta_n}$ induces a (measurable) pullback map from
$\Omega$ to $\Gamma^{\Delta_n}$ defined as
\begin{equation}
i_{\Delta_n *}[z(\cdot)] = (z(t_1), z(t_2), \ldots, z(t_n)).
\end{equation}
In its turn, we have a push-forward map $i_{\Delta_n}^*: F(\Gamma^{\Delta_n})
\rightarrow F(\Omega)$, which reads explicitly as
\begin{equation}
i^*_{\Delta_n}f [z(\cdot)] = f(z(t_1), z(t_2), \ldots, z(t_n)), \label{cylinder}
\end{equation}
where $f$ is a function on $\Gamma^{\Delta_n}$. All random variables on $\Omega$ that
can be written as a pushforward of the form (\ref{cylinder}) are called {\em cylinder
functions} with support on $\Delta_n$. They generate a large class of random variables
on $\Omega$ through a standard procedure: if $F$ is a cylinder function with support
on $\Delta_n$, we can define the $L^1$ norm
\begin{equation}
||F|| = \sum_{i=}^n |F(z(t_i))| (t_i - t_{i-1})
\end{equation}
where we take $t_0 = 0 $. One can complete the space of cylinder functions with
respect to this norm, to get a subset $B(\Omega)$ of $F(\Omega)$, consisting of
bounded functions, many of which have interesting properties (e.g. continuity).

Now, if there exists a decoherence functional on $\Omega$, it can be pullbacked on
$\Gamma^{\Delta_n} \times \Gamma^{\Delta_m}$, defining decoherence functionals
$\Phi_{\Delta_n \times \Delta_m}$ by its action on functions $f,g \in
F(\Gamma^{\Delta_n})$
\begin{equation}
\Phi_{\Delta_n \times \Delta_m} (f,g) = \Phi(i^*_{\Delta_n}f, i^*_{\Delta_m}g)
\end{equation}

Now $\Gamma$ is equipped with a Lebesque measure $dz$ (if it is a symplectic manifold
the measure is defined by the symplectic form), with respect to which we can write
$\Phi_{\Delta_n \times \Delta_m }$ in terms of distribution functions
\begin{eqnarray}
\Phi_{\Delta_n \times \Delta_m}(\prod_{t_n} dz_{t_n}, \prod_{t'_n}
dz'_{t'_n}) = \nonumber \\
\upsilon_{\Delta_n \times \Delta_m}( z_{t_1}, \ldots, z_{t_n}; z'_{t'_1}, \ldots,
z'_{t'_m}) \prod_{t_i} dz_{t_i} \prod_{t'_i} dz'_{t'_i} \label{hierarchy}
\end{eqnarray}

This implies that the full information of the decoherence functional's action on
cylinder sets (and by continuity on the whole of $B(\Omega)$) is contained in a
hierarchy of distribution functions
\begin{eqnarray}
\upsilon^{n,m}(z_1,t_1; z_2, t_2; \ldots; z_n,t_n| z'_1, t'_1; z'_2, t'_2; \ldots ;
z'_m, t'_m),
\end{eqnarray}
for all integer values of $n$ and $m$. Note that $\upsilon_{n,m}$ is here viewed as a
function on $\Gamma^{n+m} \times T^{n+m}$. These functions are {\em time-ordered} in
the sense that their temporal arguments satisfy $t_1 < t_2 < \ldots < t_n$ and $t'_1 <
t'_2 < \ldots < t'_m$.

Conversely, if we are provided with such a hierarchy of functions, we can define a
decoherence functional on $\Omega$, provided some conditions are valid.

First, for each choice of values of $\Delta_n = \{t_1, \ldots, t_n \}$ and  $\Delta_m
= \{ t'_1, \ldots, t'_m \}$, the corresponding functions on $\Gamma^{\Delta_n} \times
\Gamma^{\Delta_m}$ must be such a to define a decoherence functional  satisfying
properties C1-C6.

Second, there needs to be a compatibility condition between the decoherence
functionals in $\Delta_n \times \Delta_m$ and $\Delta'_n \times \Delta'_m$, if
$\Delta_n \subseteq \Delta'_n$ and $\Delta_m \subseteq \Delta'_m$, if both objects are
to correspond to the same decoherence functional on $\Omega \times \Omega$. For this
reason we demand the
\\ \\
 {\em (Kolmogorov) Additivity condition:}\\ $\int dz_{n+1}
\upsilon^{n+1,m}((z_1,t_1; z_2, t_2; \ldots; z_n,t_n; z_{n+1},t_{n+1}| z'_1, t'_1;
z'_2,
t'_2; \ldots ; z'_m, t'_m) =$ \\
$ \upsilon^{n,m}(z_1,t_1; z_2, t_2; \ldots; z_n,t_n| z'_1, t'_1; z'_2, t'_2; \ldots ;
z'_m, t'_m)$.
\\ \\
The satisfaction of these conditions guarantees the existence of a well-behaved
decoherence functional $\Phi$ on $B(H)$, such that its pull-backs on
$\Gamma^{\Delta_n} \times \Gamma^{\Delta_m}$ are given by the functions of the
hierarchy. ( $\Phi$ is said to be the inductive limit of this hierarchy.)

Similar hierarchies have been previously identified in the context of the phase space
picture of quantum mechanics: in reference \cite{Sri77} such hierarchies are said to
correspond to a generalisation of stochastic processes, while in reference
\cite{GHT79} they are employed to prove that the quantum mechanical predictions cannot
be recovered by any classical stochastic process.

There are other hierarchies of functions that can be used to define the decoherence
functional. For instance, given the ordered distribution functions $\upsilon^{n,m}$,
we can write the object
\begin{eqnarray}
\theta(t_n - t_{n-1}) \ldots \theta(t_2-t_1)
\theta(t'_m - t'_{m-1}) \ldots \theta(t'_2 - t'_1) \nonumber \\
\upsilon^{n,m}(z_1,t_1;  \ldots; z_i, t_i; \ldots; z_j,t_j; \ldots  z_n,t_n| z'_1,
t'_1;  \ldots ; z'_m, t'_m)
\end{eqnarray}
By fully symmetrising this object {\em separately} with respect to its $z_i,t_i$ and
$z'_i,t'_i$ entries, we obtain the {\em time-symmetric} distribution functions
$w^{n,m}$. One can obtain the decoherence functional from such a  hierarchy, only of
course now one has to demand the additional
\\ \\
{\em Symmetry postulate:}\\ $w^{n,m}(z_1,t_1;  \ldots; z_i, t_i; \ldots; z_j,t_j;
\ldots  z_n,t_n| z'_1, t'_1;  \ldots ; z'_m, t'_m) = $ \\ $w^{n,m}(z_1,t_1;  \ldots;
z_j, t_j; \ldots; z_i,t_i; \ldots  z_n,t_n| z'_1, t'_1;  \ldots ; z'_m, t'_m)$.
\\ \\
We should also remark that the requirement of Kolmogorov additivity means that we do
not need to specify the full hierarchy $\upsilon^{n,m}$ to determine the decoherence
functional. It suffices to specify the {\em diagonal} distribution functions
$\upsilon^{N,N}$ for the values of time such that $t_i = t'_i$, i.e. we just need
provide
\begin{eqnarray}
\upsilon^{N,N}(z_1,t_1;z_2,t_2; \ldots; z_N,t_N| z'_1,t_1; z'_2,t_2; \ldots ;
z_N',t_N)
\end{eqnarray}
Every distribution $\upsilon^{n,m}$ can be obtained from the $\upsilon^{N,N}$ for $N =
n+ m$. One chooses $n$ values of the time label (for simplicity take $t_1 \ldots t_N$)
 as corresponding to the forward in time entries and the remaining $m$ as corresponding to the backward in time ones (for simplicity take $t_{n+1}, \ldots t_{n+m}$). Then integrate out over $z'_1, \ldots z'_n$ and $z_{n+1} \ldots z_n$ to get the hierarchy $\upsilon^{n,m}$.
\\ \\
Summarising this section we can say that one can define a decoherence functional on
$\Omega$, by specifying a hierarchy of distribution functions (\ref{hierarchy}). We
can, therefore, study continuous-time objects, while working with discrete-time
expressions.

\subsubsection{Correlation functions}
Part of the definition of stochastic processes is the specification of a family of
functions $F_t^i$. Recalling the discussion in section 2.3, we can see that one can
readily define the mixed correlation functions $G^{n,m}$ as
\begin{eqnarray}
G^{n,m}(a_1,t_1; a_2,t_2; \ldots ; a_n,t_n| b_1,t'_1; b_2,t'_2; \ldots ; b_m,t'_m) =
\\ \nonumber
 \Phi(F^{a_1}_{t_1} F^{a_2}_{t_2}
\ldots F^{a_n}_{t_n}, F^{b_1}_{t'_1} F^{b_2}_{t'_2} \ldots F^{b_m}_{t'_m})
\end{eqnarray}

Note that each correlation function $G^{n,m}$ (for fixed values of $n$ and $m$) needs
only the information contained in $\upsilon^{n,m}$ of equation (\ref{hierarchy}) in
order to be determined.

As we discussed in section 2. $G^{n,0}$ are the time-ordered correlation functions,
$G^{0,m}$ the anti-time-ordered correlation functions. In general there are certain
relations between functions with the same value of $n+m$. For instance
\begin{equation}
G^{2,0}(a_1,t_1;a_2,t_2) = \theta(t_2- t_1) G^{1,1}(a_1,t_1|a_2,t_2) + \theta(t_1-t_2)
G^{1,1}(a_2,t_2|a_1,t_1)
\end{equation}

Once we have the hierarchy $G_{n,m}$ of correlation functions associated to $F^a_t$ we
can define the corresponding closed-time-path generating functional $Z_F[J_+,J_-]$,
which is written in terms of the sources $J^{a}_+(t), J^{a}_-(t)$ as
\begin{eqnarray}
Z_F[J_+,J_-] = \sum_{n=0}^{\infty} \sum_{m=0}^{\infty} \frac{i^n (-i)^m}{n! m!}
\hspace{5cm}
 \nonumber \\ \times \sum_{a_1,\ldots a_n} \sum_{b_1,
\ldots, b_m}  \int dt_1 \ldots dt_n dt'_1 \ldots dt'_m \hspace{3.5cm}
\nonumber \\
\times G^{n,m}(a_1,t_1; \ldots ; a_n,t_n| b_1, t'_1; \ldots ; b_m, t'_m) \hspace{2cm}
 \nonumber \\
\times J^{a_1}_+(t_1) \ldots J^{a_n}_+(t_n) J^{b_1}_-(t'_1) \ldots
 J^{b_m}_-(t'_m)
\end{eqnarray}

Clearly the CTP generating functional can be written as
\begin{equation}
Z_F[J_+,J_-] = \Phi(e^{i F \cdot J_+}, e^{-i F \cdot J_-}),
\end{equation}
where $F \cdot J_{\pm} : = \int dt \sum_i F^{i}_t J_{\pm}î(t)$. From this equation it
is easy to see the two main properties that $Z_F[J_+,J_-]$ inherits from the
decoherence functional:
\\ \\ \\
1. {\em Hermiticity:} $Z_F[J_-,J_+] = Z^*_F[J_+,J_-]$. \\ \\
2. {\em Normalisation:} $Z_F[0,0] = 1$. \\ \\
\\
 In an inverse way one can obtain the correlation functions
from $Z[J_+,J_-]$ by functional differentiation
\begin{eqnarray}
G_F^{n,m} (a_1,t_1; \ldots ; a_n , t_n | b_1 t'_1; \ldots,b_m, t'_m) = \nonumber \\
 (-i)^n i^m \frac{\delta^n}{\delta J^{a_1}_+(t_1) \ldots \delta
J^{a_n}_+(t_n)} \frac{\delta^m}{\delta J^{b_1}_-(t_1) \ldots \delta J^{b_m}_-(t_m)}
Z[J_+,J_-]|_{J_+=J_-=0}.
\end{eqnarray}

If the family of functions $F^a_t$ separates $\Omega$, then the information in $Z_F$
is sufficient to reconstruct the whole decoherence functional. Such is the case, for
instance, when the $F^a_t$ are functions of type (\ref{canhis}), with $f^a$ being a
set of coordinates on $\Gamma$.

\paragraph{Gaussian processes.} As in classical probability theory
of importance are the Gaussian processes; these are processes, whose CTP generating
functional is a Gaussian function of the currents. In Gaussian processes all
correlation functions are determined by the two-point functions. Namely, if we define
\begin{eqnarray}
F^a(t) = G^{1,0}(a,t) = G^{0,1}(a,t) \\
i \Delta^{ab}(t,t') = G^{2,0}(a, t; b, t') - F^a(t) F^b(t') \\
i K^{ab}(t,t') = G^{1,1}(a,t|b,t') + G^{1,1}(b,t'|a,t) - 2 F^a(t) F^b(t'),
\end{eqnarray}
the CTP generating functional for a Gaussian process reads
\begin{eqnarray}
Z[J_+,J_-] = \exp \left( - \frac{i}{2} J_+ \cdot \Delta \cdot J_+ + \frac{i}{2} J_-
\cdot \bar{\Delta} \cdot J_- \right.
\nonumber \\
\left. + i J_+ \cdot K \cdot J_- + i (J_+ - J_- ) \cdot X \right) \label{Gauss}
\end{eqnarray}

Here   we wrote  $J \cdot \Delta \cdot J' = \int dt dt' J^a(t) \Delta^{ab}(t,t')
J^b(t')$  and  the bar is used to denote complex conjugation.

\subsection{The kinematical process}

So far our discussion of quantum processes was rather formal, that is we did not
attempt to write down processes that reproduce the quantum mechanical formalism.

In this section, we shall deal with a class of processes that is of greatest
importance for the implementation of our programme: {\em the kinematical processes},
i.e. quantum processes for systems that have vanishing quantum mechanical Hamiltonian.
Unlike classical probability theory, where the absence of dynamics makes a stochastic
process trivial, in quantum processes the full wealth of quantum mechanical behaviour
is manifested already at the kinematical level. In fact, as we shall see the
contribution of dynamics is insignificant (with respect to the defining features of
quantum processes) compared to the kinematics.

This remark goes back to Heisenberg \cite{Heis}, who actually started his
investigations of quantum theory by postulating that the difference between classical
and quantum theory is to be found at the kinematical level, in the nature of the basic
observables. In the context of histories, Savvidou \cite{Sav99a} has uncovered a sharp
distinction between kinematics and dynamics; there generically exist different groups
implementing kinematical and dynamical transformations and in history theories they
coexist. This duality between kinematics and dynamics
 corresponds to different ways times is manifested in quantum theory as "reduction
 of the wave packet" and Heisenberg dynamics respectively \cite{Sav99a}. This
remarkable property of histories is also present in the distinction between geometric
and dynamical phase \cite{AnSav02}.  These facts have enabled us to argue  in
\cite{AnSav02}, that all defining properties of quantum theory are found at solely the
{\em kinematical} level, dynamics being structurally identical with classical ones. In
this paper, we shall make this claim more concrete.

\subsubsection{Coherent states}
In our previous work \cite{An01a,An01b}, when we wanted to discuss the definition of a
process in phase space that mimic quantum phenomena, we instictively assumed that the
natural way to do so was through the use of the Wigner transform. In other words, we
obtained the functions $\upsilon_{n,m}$, that define the decoherence functional on
phase space by Wigner-transforming the corresponding objects for the Hilbert space
decoherence functional. One can then obtain a quantum process on the phase space,
through the construction outlined in section 4.2. The expressions for $\upsilon_{n,m}$
thus obtained were rather unwieldy and were not useful in proceeding further.

We found that there is a dramatic simplification if we define the phase space
distribution functions by means of the coherent states. Recall, that the coherent
states are, in general, associated with a canonical group, that is a group $G$ that
has an irreducible representation on the Hilbert space $H$, by unitary operators
$\hat{U}(g)$ \cite{KlSk85, Per86}. Taking a reference normalised vector $| 0 \rangle$
(often a vacuum state) we can define the Hilbert space vectors $\hat{U}(g)|0 \rangle$.
Now we define the equivalence relation
\\ \\
$g \sim g'$ if $\hat{U}(g) |0 \rangle = e^{i \phi} \hat{U}(g') | 0 \rangle$, for some
phase $e^{i \phi}$.
\\ \\
The quotient space $\Gamma = G/ \sim$ is the parameter space labeling the coherent
states $| z \rangle, z \in \Gamma$. $\Gamma$ has a rich structure, since the coherent
states define an injection map $i$ from $\Gamma$ to the projective Hilbert space $PH$.
The latter has a Riemannian metric $ds^2$ (coming from the real part of the Hilbert
space's inner product), a symplectic form $\omega$ coming from the imaginary part of
the inner product, the Hopf bundle (which we discussed in section 2.2) and a $U(1)$
connection (the Berry connection $A$), such that $\omega = d A$. All these structures
can be pullbacked through $i$ on $\Gamma$ making it a symplectic manifold, with a
$U(1)$ bundle, a $U(1)$ connection one-form $A$ and a Riemannian metric. Explicitly,
these structures read
\begin{eqnarray}
ds^2(\Gamma) &=& ||d |z \rangle ||^2 - |\langle z|d |  z \rangle |^2
\\
A &=& -i \langle z|d| z \rangle \\
\omega &=& dA,
\end{eqnarray}
where $d$ is the exterior derivative operator on $\Gamma$.

Also $\Gamma$ is a homogeneous space as the group $G$ acts transitively on $\Gamma$
through the map $|z \rangle \rightarrow \hat{U}(g) | z \rangle$.

\subsubsection{Defining the kinematical process}

Let us consider a Hilbert space $H$ carrying a representation $\hat{U}(g)$ of the
group $G$, and a choice of reference vector $| 0 \rangle$ leading to a family of
coherent states $| z \rangle, z \in \Gamma $. If we denote by $\hat{A}^a$ the
self-adjoint operators that generate the group elements, we can define the basic
functions $f^a$ on $\Gamma$ through
\begin{equation}
\hat{A}^a = \int dz f^a(z) |z \rangle \langle z |. \label{Psymbol1}
\end{equation}
The functions $f^a$ are known as the $P$-symbols of $\hat{A}^a$ and we implicitly
restrict to operators, whose $P$-symbols are measurable functions.

Now, having $\Gamma$ one can define the space $\Omega$ of continuous paths on $\Gamma$
and the family of functions $F^a_t$ associated to the $f^a$ of equation
(\ref{canhis}). All is missing from the definition of  a quantum processes is the
specification of a decoherence functional. This is achieved by specifying the
hierarchy of ordered distribution functions $\upsilon^{n,m}$. To do so, we write the
time instants in terms of their ordering $t_1 \leq t_2 \leq \ldots t_n$, and $t'_1
\leq t'_2 \leq \ldots \leq t'_m$. If we  write $\hat{\alpha}_z = | z \rangle \langle z
|$ we will have
\begin{eqnarray}
\upsilon^{n,m}_{z_0}(z_1,t_1; z_2,t_2; \ldots ; z_n, t_n|z'_1,t'_1;
z'_2, t'_2; \ldots ; z'_m,t'_m) = \nonumber \\
Tr    \left( \hat{\alpha}_{z_n} \hat{\alpha}_{z_{n-1}} \ldots \hat{\alpha}_{z_2}
\hat{\alpha}_{z_1} \hat{\alpha}_{z_0} \hat{\alpha}_{z'_1} \ldots
\hat{\alpha}_{z'_{m-1}}
\hat{\alpha}_{z'_m} \right) = \nonumber \\
\langle z'_m| z_n \rangle \langle z_n| z_{n-1} \rangle \ldots \langle z_2| z_1 \rangle
\langle z_1| z_0 \rangle \langle z_0| z'_1 \rangle \langle z'_1| z'_2 \rangle \ldots
\langle z'_{m-1}| z'_m \rangle .
\end{eqnarray}

This process is defined as starting from $z_0 \in \Gamma$ at $t = 0$. One can define
more general processes by smearing through $z_0$ with a distribution function positive
$\rho(z_0)$
\begin{equation}
\upsilon^{n,m}_{\rho} = \int dz_0 \rho(z_0) \upsilon^{n,m}_{z_0}.
\end{equation}
In the Hilbert space language, this amounts to an initial density matrix
\begin{equation}
\hat{\rho} = \int dz \rho(z) |z \rangle \langle z |
\end{equation}
\\ \\
Let us note some distinguishing features of the kinematic process:
\\ \\ \\
1. The expression for the distribution function factorises in products of the form
$\langle z| z' \rangle$. The knowledge of this inner product, suffices to fully
determine the kinematical process. In fact, the distribution function $\upsilon^{n,m}$
is known as the $n+m+1$ Bargmann invariant \cite{MuSi93}.
\\ \\
 2. The distributions $\upsilon^{n,m}$ do not depend on the
values of time $t$, only on their ordering. The same is true for $t'$. More than that,
if we consider the following cyclic ordering for the time instants $t_0 \rightarrow
t_1 \rightarrow t_2 \rightarrow \ldots \rightarrow t_n \rightarrow t'_m \rightarrow
\ldots \rightarrow t'_2 \rightarrow t'_1 \rightarrow t_0$, the distributions are
invariant if we consider any time as origin and then proceed cyclically along the
arrows. In other words, the kinematic process manifests the symmetry of a {\em closed
time path}.
\\ \\
3. Let us consider that the process being defined in the time interval $[0,\tau]$ and
consider the  distribution function $\upsilon^{n,m}$ for large values of $n$ and $m$.
Take for simplicity $n = m = N$. Choose also the time  instants  such that $|t_i -
t_{i-1}| \leq \delta t = \tau/N$ for all $i$ and similarly for $t'$. Also, let $t_n =
t'_m = \tau$. Then we have a discretised approximation to a decoherence functional for
continuous paths $z(\cdot), z'(\cdot)$, which  for $N \rightarrow \infty$ would
converge to
\begin{equation}
\Phi(z(\cdot),z'(\cdot)) = e^{- i \int_{C}  \langle z|d|z \rangle } + O(\delta t^2) =
e^{i \int_C A} + O(\delta t^2),
\end{equation}
where $C$ is the closed path obtained by appending the path $z'(\cdot)$ with reverse
orientation at the end of $z(\cdot)$. The distribution function for the decoherence
functional then converges at the large $N$ limit to the holonomy of the pull-back of
the Berry connection on $\Gamma$.  Of course, this convergence is to be interpreted
with a grain of salt as the  support of the decoherence functional is on cylinder
functions, rather than differentiable ones, for which the holonomy is rigorously
defined.

\subsubsection{The one-dimensional particle process}
Let us study the simplest example of a kinematical process, that of a particle at a
line. The functions determining the quantum process are the position $x_t$ and the
momentum $p_t$. Let us take for simplicity that the coherent states are Gaussian (the
overlap $\langle z|z' \rangle$ is Gaussian as in the standard case) and that $z_0 =
0$. (The initial point does not really matter as $\Gamma$ is a homogeneous space and
there is no dynamics to differentiate between points). We have the following
correlation functions
\begin{eqnarray}
G^{2,0}(x,t;x',t') = G^{0,2}(x,t;x',t') = G^{1,1}(x,t|x',t') = \sigma_x^2 \\
G^{0,2}(p,t;p',t') = G^{0,2}(p,t;p',t') = G^{1,1}(p,t|p',t') =
\sigma_p^2 \\
G^{2,0}(x,t;p,t') = G^{0,2}(p,t;x,t') = \frac{1}{2} [\theta(t-t')
(C+i) + \theta(t'-t) (C -i ) ] \\
G^{2,0}(p,t;x,t') = G^{0,2}(x,t;p,t') = \frac{1}{2} [\theta(t'-t)
(C+i) + \theta(t-t') (C -i ) ] \\
G^{1,1}(x,t|p,t') = \frac{1}{2}(C-i) = \bar{G}^{1,1} (p,t|x,t'),
\end{eqnarray}
in terms of $\sigma_x^2 = \langle 0| \hat{x}^2|0 \rangle$, $\sigma_p^2 = \langle 0 |
\hat{p}^2|0 \rangle $ and $C = \langle 0 | \hat{x} \hat{p}+ \hat{p} \hat{x}| 0
\rangle$.

The corresponding CTP generating functional is of the form (\ref{Gauss}) with kernels
\begin{eqnarray}
i \Delta(t,t') &=& \left( \begin{array}{cc} \sigma_x^2 & \frac{1}{2}
[C + i \eta(t-t')] \\
\frac{1}{2} [C - i \eta(t-t') ]&
\sigma_p^2 \end{array} \right) \nonumber \\
i K(t,t') &=& \left( \begin{array}{cc} \sigma_x^2 & \frac{1}{2}(C-i)
\\ \frac{1}{2} (C+i) & \sigma_p^2 \end{array} \right),
\end{eqnarray}
where $\eta(t-t') = \theta(t-t') - \theta(t'-t)$.

 The fact to note is that it is the correlation between
position and momentum that causes the CTP generating functional to be complex. In
absence of this the kinematical process would have a completely real-valued generating
functional.

\subsection{Introducing dynamics}

The standard way to introduce dynamics is by the introduction of a Hamiltonian
operator. Its effect is a change and the introduction of an explicit time dependence
in the kernels $\langle z|z' \rangle$. In other words, the definition proceeds
similarly as the kinematical process, only now the distributions $\upsilon^{n,m}$ read
\begin{eqnarray}
\upsilon^{n,m}_{z_0}(z_1,t_1; z_2,t_2; \ldots ; z_n, t_n|z'_1,t'_1;
z'_2, t'_2; \ldots ; z'_m,t'_m) =  \nonumber \\
\langle z'_m|e^{-i \hat{H}(t'_m - t_n)} |z_n \rangle \langle z_n| e^{-i
\hat{H}(t_{n}-t_{n-1})}| z_{n-1} \rangle \ldots
\nonumber \\
\times \langle z_2|e^{-i \hat{H}(t_2-t_1)}| z_1 \rangle \langle z_1|e^{-i
\hat{H}(t_1-t_0)}| z_0 \rangle \nonumber \\
\times \langle z_0| e^{i \hat{H}(t'_1-t_0)}| z'_1 \rangle \langle
z'_1|e^{i\hat{H}(t'_2-t'_1)} | z'_2 \rangle \ldots \langle z'_{n-1}|e^{i
\hat{H}(t'_n-t'_{n-1})} |z'_n \rangle .
\end{eqnarray}

We see then that the basic object by which the decoherence functional is constructed
is the "coherent state propagator" $ \chi(z,z'|t): =  \langle z|e^{-i \hat{H}t}|z'
\rangle $.

Now, let us consider the following. Consider a process with respect to the {\em
kinematical } decoherence functional (which we shall denote as $\Phi_0$), but which is
defined with respect to another set of variables $^H\!F^a_t$, defined as
\begin{equation}
\hat{A}^a(t) = \int dz  {}^H\!F_t^a[z(\cdot)] |z(t) \rangle \langle z(t)|
\end{equation}
in terms of  the Heisenberg picture operators
\begin{equation}
 \hat{A}^a(t) = e^{i\hat{H}(t-t_0)} \hat{A}^a e^{-i \hat{H}(t
-t_0)}.
\end{equation}
  Note that the construction of these objects employ the
time label $t$ in two distinct fashions, one kinematical as the argument of the
coherent state and one dynamical in the exponentiation of the Hamiltonian.

The important point is that the process $(\Omega, \Phi_H, F^a_t)$ is completely
isomorphic to the process $(\Omega, \Phi_0, {}^HF^a_t)$ as they have completely
identical correlation functions.

The physical interpretation of this fact has been given by Savvidou \cite{Sav99a}. Let
us define $X^a_t(s)$ by
\begin{equation}
e^{i \hat{H}s} \hat{A}^a e^{-i\hat{H}s} = \int dz_t X^a_t(s) |z_t \rangle \langle z_t|
z(t) \rangle .
\end{equation}
It is true that $X^a_t(s = t-t_0) = ^H\!F_t$. $X^a_t (s)$ can be obtained from $F^a_t$
by the solution of a deterministic set of equations. To see this note that for small
$s$
\begin{equation}
e^{i \hat{H}s} \hat{A}^a e^{-i\hat{H}s} \simeq {A}^a + i s \int dz_t X^a_t(z_t)
[\hat{H}, \hat{P}_{z_t}],
\end{equation}
where $\hat{P}_{z} = |z \rangle \langle z|$. If we define the kernel
$\alpha(z,z',z'')$ by
\begin{equation}
[\hat{P}_z,\hat{P}_z'] = \int dz'' \alpha(z,z',z'') \hat{P}_{z''},
\end{equation}
and write $\hat{H} = \int dz h(z) \hat{P}_z$, in terms of the $P$-symbol $h$ of the
Hamiltonian,  it is easy to verify that

\begin{equation}
\frac{d}{ds} X^a_t(s) = V_H(X^a_t), \label{VH}
\end{equation}
where $V_H$ is an operator on functions on $\Gamma$ givn by

\begin{equation}
V_H(f)(z) = \int dz' \beta(z,z') f(z'),
\end{equation}
in terms of $\beta(z,z') = \int z'' \alpha(z',z'',z) h(z'')$.

The initial conditions fpr these equations are $X^a_t(s = t_0) = F^a_t$. This means
that the process $(\Omega, \Phi_H, F^a_t)$ is obtained by first evolving the $F^a_t$
according to deterministic equations of motion - separately for each $t$- and then
constructing the correlation functions, when the initial conditions are distributed
according to the {\em kinematic process}.

\subsection{Quantum differential equations}

We can make the statement  about the overriding importance of kinematics more precise,
by introducing the notion of a {\em quantum differential equation}. This is meant to
be the analogue of a stochastic differential equation in classical probability theory.

Such equations arise when  one seeks to understand, how single time observables, i.e.
functions on $\Gamma$ change in time. Let us consider an observable $f^a$ that has
some fixed value at time $t$ as the system is in the state $z(t)$. In fact, we shall
have that
\begin{equation}
f^a(z(t)) = F^a_t[z(\cdot)]
\end{equation}
for the process $F^a$. Let us consider the description of the system  in terms of the
process $(\Omega, \Phi_0, ^HF_t^a)$. At the next moment $t + \epsilon$, $f^a$ will go
the function $\tilde{f}^a = f^a + \epsilon V_H(f^a) +O(\epsilon^2)$, which is however
a function on $\Gamma_{t+\epsilon}$. This implies that
\begin{eqnarray}
\delta f^a (z(t))) := \tilde{f}^a(z(t+\epsilon))  - f^a(z(t))  \nonumber \\
= f^a(z(t + \epsilon)) - f^a(z(t))+  \epsilon V_H(f^a)(z(t)) +O(\epsilon^2) \nonumber \\
= (F^a_{t+\epsilon} - F^a_t) + \epsilon V_H(f^a)(z(t))
\end{eqnarray}

Taking $\epsilon = \delta t$ we have the formal equation
\begin{equation}
d f^a(t) - V_H(f^a(t)) = dF_t^a ,
\end{equation}
which can also be written as
\begin{equation}
\frac{df^a}{dt} - V_H(f^a(t)) = \dot{F}_t^a \label{quantdiffeq}
\end{equation}

 This equation is the quantum analogue of a stochastic differential equation.
It can be interpreted as referring to  {\em an individual quantum system} and stating
that the rate of change of any function $f$ equals a deterministic part plus a random
``external force'', which is distributed over the ensemble according to the {\em
kinematical process } of the system. As we argued earlier, quantum theory has its
origins in the kinematical process.

\paragraph{The particle at a line} Let us now consider the
correlation functions  $\dot{F}^a_t$ for the particle in one dimension. It is easy to
check that the expectation values for $\dot{x}$ and $\dot{p}$ vanish in the
kinematical process and so do all two-point correlation functions that contain a pair
of $\dot{x}$ and ${\dot p}$ 's. We have however,
\begin{eqnarray}
G^{2,0}(\dot{x},t;\dot{p},t') = G^{0,2}(\dot{p},t;\dot{x},t') =
-\frac{i}{2} \partial_t^2 \eta(t-t') \\
G^{2,0}(\dot{p},t;\dot{x},t') = G^{0,2}(\dot{x},t;\dot{p},t') =
\frac{i}{2} \partial_t^2 \eta(t-t') \\
G^{1,1}(x,t|p,t') = G^{1,1}(p,t|x,t') = 0.
\end{eqnarray}
By $\dot{x}$ we denote $\frac{1}{\epsilon} (x_{t+ \epsilon} - x_t)$ for positive
$\epsilon \rightarrow 0$  and similarly for $\dot{p}$. While $\partial_t^2 \eta(t-t')$
denotes the limit of $\eta(t-t') + \eta(t-t'- \epsilon) - 2\eta(t-t')$, in terms of
$\eta(t-t')$. This is a version of the first derivative of the $\delta$-function.

 The corresponding CTP generating functional is
 \begin{equation}
Z_{CTP}[J_+,J_-] = \exp \left(- \frac{i}{4}J^x_+ \cdot \dot{J}_+^p + \frac{i}{4} J^x_-
\cdot \dot{J}_-^p \right),
 \end{equation}
where $J^x_{\pm}$ and $J^p_{\pm}$ are the source terms for $\dot{x}$ and $\dot{p}$
respectively.

It is interesting to write the quantum differential equation for a particle with a
Hamiltonian $\hat{H} = \frac{\hat{p}^2}{2m} + V(\hat{x})$. If we denote as $\xi^x_t =
\dot{X}_t$ and $\xi^p_t = \dot{P}_t$  the external sources associated with the
kinematic process, the quantum differential equations read
\begin{eqnarray}
\dot{x} - \frac{1}{m} p = \xi^x_t \nonumber \\
\dot{p} + V'(x) = \xi^p_t,
\end{eqnarray}
which implies that
\begin{equation}
\ddot{x} + V'(x) = \Xi_t,
\end{equation}
where
\begin{equation}
\Xi_t = \dot{\xi}^x_t - \frac{1}{m}\xi^p_t.
\end{equation}
This implies that the external force for the configuration space equation of motion is
a measure of the failure of the momentum to coincide with particle velocity in the
kinematic process, something anticipated in \cite{Sav99b}. It is interesting to also
remark that if we consider the case of fields, we can still write the analogue of the
quantum differential equation on configuration space. For a scalar field this would be
\begin{equation}
 \ddot{\phi} - \nabla^2 \phi - m^2 \phi - V'(\phi) = \Xi_t
\end{equation}
While the left-hand side is the deterministic Lorentz invariant equation of motion,
the right-hand side, which contains the quantum mechanical contribution,  explicitly
depends on the choice of time variable through the choice of momentum  and hence the
quantum differential equation {\em does depend on the choice of the spacetime
foliation}. This fact is not apparent (but still present) in standard quantum theory,
has been identified by Savvidou in the context of continuous-time histories
\cite{Sav01a}.

Before concluding this section, we want to remark on the appealing possibility that
equation (4.43) can be interpreted as referring to an individual system in analogy to
the classical Langevin equations. That is, we can consider that equation
(\ref{quantdiffeq}) refers to an individual system (a particle), which is found within
a ``fluctuating environment'', that induces the ``random forces'' $\xi^a(t) =
\dot{F}^a_t$. However, these forces are not distributed according to a classical
probability distribution, but according to the kinematic processes (and are possibly
geometrical in origin).

We are not  in a position to argue, whether this interpretation should be taken
seriously or not. The reasons are partly mathematical and partly physical: from the
mathematical side we need to verify that equations such as (\ref{quantdiffeq}) are
more than empty symbols: is it actually a type of equation that can admit solutions?
we hope to justify such equations by adopting the theory of stochastic integrals (of
Ito) in the quantum context. From a physical point of view, even though we are
committed to finding a description for the individual quantum system, the picture of a
particle moving under random forces is not necessarily our first choice: it is perhaps
too classical. Nonetheless, equation (\ref{quantdiffeq}) has large theoretical
interest and we would like to see, whether it would be possible to simulate its
solutions  numerically as we can do with stochastic processes. This would provide  a
way of generating actual trajectories for individual quantum systems.

\section{Recovering quantum mechanics}
In the previous section, we showed how to obtain quantum processes starting from
quantum theory. Now, we want to invert this procedure and ask how one can obtain
standard quantum theory starting from a generic quantum process, that satisfies the
axioms stated in section 4.1.

\subsection{The Markov property for quantum processes}
One important feature of quantum dynamics is that the time evolution of the object
encoding the probabilities (the wave function or the density matrix) is given by a
linear partial differential equation; this means that the knowledge of this object at
a moment of time suffices to determine it at any future instant.

The analogous property for classical processes  is known as the Markov property.  We
shall here try to study in more detail its quantum analogue.

\subsubsection{Conditioning with respect to past}

Let us consider a quantum process $(\Omega,\Phi, F^a_t)$,  where we choose the $F^a_t$
to correspond to a complete set of coordinate functions on the single-time manifold
$\Gamma$. We will denote $F_t: \Omega \rightarrow {\bf R}^n$.

We can define two subfields. One is the {\em instant} subfield ${\cal A}_t$, which is
generated by all sets of the  form $(F_t)^{-1}(B)$, for a fixed value of $t$ and any
Borel subset $B$ of ${\bf R}^n$. The other is the {\em past } subfield ${\cal A}_{\leq
t}$, which is generated by all sets of the form $(F_s)^{-1}(B)$. for $ t_0 \leq s \leq
t$ and Borel subsets $B$ of ${\bf R}^n$.

The {\em Markov property} is phrased in terms of the  conditional pairs. If we have
the times $t,t' \geq s \geq t_0$ , then a process $(\Omega,\Phi, F^a_t)$ satisfies the
Markov property if
\begin{equation}
\tilde{\Phi}((F_t)^{-1}(C),(F_{t'})^{-1}(C')|{\cal A}_s) =
 \tilde{\Phi}((F_t)^{-1}(C),(F_{t'})^{-1}(C')|{\cal A}_{\leq s}),
\label{Markov}
\end{equation}
for any Borel sets $C$ and $C'$ in ${\bf R}^n$.

This states that once we have the full knowledge of the decoherence functional at a
moment $t$, we need no knowledge from its past in order to define probabilities and
phases for any future measurements.

\subsubsection{The quantum Chapman-Kolmogorov equation}

Now $\tilde{\Phi}((F_t)^{-1}(C),(F_{t'})^{-1}(C')|{\cal A}_s)$ is a functional on
${\bf R}^n \times {\bf R}^m$, and can  be written in terms of the Lebesque measure
$d^nz_t$ in ${\bf R}^m$, which induced through the coordinate functions $F_t^a$ from
the measure $dz$ on $\Omega$. Explicitly we would have
\begin{equation}
\tilde{\Phi}((F_t)^{-1}(C),(F_{t'})^{-1}(C')|{\cal A}_s) = \int \chi_C(z)
\chi_{C'}(z')\upsilon(z,t;z',t'|{\cal A}_s) d^nz_t d^nz_{t'} \label{propagfirst}
\end{equation}

Now, we have
\begin{eqnarray}
\tilde{\Phi}((F_t)^{-1}(C),(F_{t'})^{-1}(C')|{\cal A}_s) =
\tilde{\Phi}(\chi_C(F_t), \chi_{C'}(F_{t'})|{\cal A}_s) \nonumber \\
=   \tilde{\Phi}(\chi_C(F_t), \chi_{C'}(F_{t'})|{\cal A}_{\leq s}) =
\tilde{\Phi}[\tilde{\Phi}(\chi_C(F_t), \chi_{C'}(F_{t'})|{\cal A}_{\leq s})|{\cal
A}_{{\leq s'}}], \label{Mstep1}
\end{eqnarray}
where $s' \geq s$, by  virtue of (\ref{condtwo}) and since obviously ${\cal A}_{\leq
s} \subseteq {\cal A}_{\leq s'}$ if  $s \leq s'$. (Note
 the use of the Markov property in equation (\ref{Mstep1}).)

We can further work on this equation to get
\begin{eqnarray}
\tilde{\Phi}((F_t)^{-1}(C),(F_{t'})^{-1}(C')|{\cal A}_s)=
\tilde{\Phi}[\tilde{\Phi}(\chi_C(F_t), \chi_{C'}(F_{t'})|{\cal A}_{\leq s'})|{\cal
A}_{\leq s}]
 \nonumber \\
= \tilde{\Phi}[\tilde{\Phi}(\chi_C(F_t), \chi_{C'}(F_{t'})|{\cal A}_{ s'})|{\cal
A}_{s}] \label{Mstep2}
\end{eqnarray}

Now let us recall that $\tilde{\Phi}(\cdot,\cdot|{\cal A}_s) =
\tilde{\Phi}(\cdot,\cdot|F_s)$ is a function on $\Omega \times \Omega$. Since $F_s$
are coordinate functions and locally in one-to-one correspondence with points of
$\Gamma_s$ one can write this as
\begin{equation}
\tilde{\Phi}(\cdot,\cdot|z,z',s)
\end{equation}
Hence, the corresponding densities $\upsilon (z, t; z',t'|{\cal A}_s)$  can be
represented by a kernel with arguments on $\Gamma$ as $
\upsilon(z_1,t;z_1',t'|z_0,z'_0,s)$. Taking $t=t'$ we can define the  function
\begin{equation}
\upsilon(z_1,z_1';t|z_0,z'_0;s),
\end{equation}
which is essentially the {\em density matrix propagator}.

If we write $\tilde{\Phi}$ in terms of this kernel in equation (\ref{Mstep2}) we will
obtain
\begin{equation}
\upsilon(z_1,z_1';t|z_0,z'_0;s) = \int dz dz' \upsilon(z_1,z_1';t|z,z';s')
\upsilon(z,z';s'|z_0,z'_0; s) \label{ChapKol}
\end{equation}
This  property of Markov processes is a quantum version of  the Chapman-Kolmogorov
equation.

The quantum Chapman-Kolmogorov equations implies, that the knowledge of the propagator
$\upsilon$ together with the decoherence functional at the initial moment of time
suffices to determine the full hierarchy of distribution functions.

Indeed, if we get a density  $\rho_0(z_0,z_0') $ for time $t=0$, one can determine the
diagonal distribution functions
\begin{eqnarray}
\upsilon^{N+1,N+1}(z_0,t_0;z_1,t_1; \ldots ; z_{N},t_{N-1}|z'_0,t_0; z_1,t_1;
 \ldots ; z'_{N-1},t_{N-1}) =
\nonumber \\
\upsilon(z_{N},z'_{N};t_{N}|z_{N-1},z'_{N-1};t_{N-1})
 \ldots \upsilon(z_1,z'_1;t_1|z_0,z'_0;t_0)
\rho_0(z_0,z'_0). \label{factorisation}
\end{eqnarray}

As we explained in section 4.1 these distribution functions contain  enough
information to determine the hierarchy $\upsilon^{n,m}$ and hence the decoherence
functional.

\subsubsection{Symmetries of the propagator}

By virtue of its definition and the properties of the decoherence functional
 it is easy to demonstrate that the propagator satisfies the
 following properties \\ \\
{\em 1. Hermiticity:}
$\upsilon(z,z';t|z_0,z'_0;s) = \bar{\upsilon} (z',z;t|z'_0,z_0;s).$ \\
This is clearly a consequence of the hermiticity property of the decoherence
functional and is inherited into the conditional pair,
by which the propagator is defined. \\ \\
{\em 2. Normalisation:} $ \int dz dz' \upsilon(z,z';t|z_0,z'_0;s)
= 1$. \\
This follows from the normalisation condition, by virtue of (\ref{condnorm}) .
\\ \\
{\em 3. Positivity:} If $\rho_0$ is  positive, then so is $\int dz_0 dz'_0
\upsilon(z,z';t|z_0,z'_0;s) \rho_0(z_0,z'_0)$. A function $\rho(z,z')$ is called
positive, if it satisfies the positivity condition of the decoherence functional, i.e.
for any (complex-valued) function $F$,
\begin{equation}
\int dz dz' \rho(z,z') F(z) F^*(z') \geq 0.
\end{equation}
It is clear that  this condition  follows from the positivity property of the
decoherence functional.
\\
\\
These properties of the decoherence functional can be written in an operator language.
For this purpose, let us  consider the Hilbert space $ V = {\cal L}^2(\Gamma,dz)$ and
its dual $\bar{V}$. This is {\em not} the physical Hilbert space of the corresponding
quantum theory; it is simply introduced as a  convenient way to describe the
conditions 1-3 for the propagator.

 We shall  denote elements of $V$ as
$f_{\rightarrow}$, of $\bar{V}$ as $g_{\leftarrow}$ and as $g_{\leftarrow} \cdot
f_{\rightarrow}$ the operation of the natural pairing between $V$ and $\bar{V}$. We
employ the arrows in order to keep track in which Hilbert space each element belongs.

In $V \otimes \bar{V}$ we have the vectors $|fg) = f_{\rightarrow} \otimes
g_{\leftarrow}$, which have as inner product
\begin{equation}
(f'g'|fg) = (f'_{\rightarrow} \cdot f_{\leftarrow})(g_{\rightarrow} \cdot
g'_{\leftarrow})
\end{equation}

For fixed values of $t$ and $s$,  $\upsilon$ defines an operator $Y$ on $V \otimes
\bar{V}$, such that
\begin{equation}
(f'g'|Y|fg) = \int dz dz' dz_0 dz_0' \bar{f}'(z) g'(z') \upsilon(z,z';t|z_0,z_0';s)
f(z_0) \bar{g}(z_0'),
\end{equation}
for functions $f,g,f',g'$ that are elements of $V$.

 Now the pairing $\cdot$ can be
extended by linearity to a linear map from $V \otimes \bar{V}$ to
 ${\bf C}$. Given the fact that $V \otimes \bar{V}$ is a
subspace of the space of bounded linear operators $B(V)$ on $V$, the pairing $\cdot$
is identical  to the trace functional on $B(V)$. The normalisation condition  implies
that $Y$ preserves the map $\cdot$, or in other words if $Y$ is viewed as an operator
on $B(V)$ is trace-preserving.

 Now $V \otimes \bar{V}$ is the space of Hilbert-Schmidt operators on
$H$ \footnote{A Hilbert-Schmidt operator is one that has  finite value of the norm
$||A||_2 = [Tr(A^{\dagger}A)]^{1/2}$.}
 and if
$O$ is such an operator $Y$ will act  in such a way as
\begin{equation}
Tr Y(O) = Tr O.
\end{equation}
This relation is generalised to further operators besides the Hilbert-Schmidt ones by
linearity and taking limits with respect to the trace-norm topology.

The hermiticity condition, when applied to such an operator $Y$ implies that
\begin{equation}
Tr(A^{\dagger}Y(B)) = Tr[A Y(B^{\dagger})],
\end{equation}
for Hilbert-Schmidt operators $A$ and $B$.

 The positivity condition is translated to the fact, that if $O$
is a positive operator then $Y(O)$ is also positive.

One way to see the physical meaning of these conditions is to consider the class of
trace-preserving operators of the form $Y(O) = \sum_i c_i A_iO A^{-1}_i$, for some
operators $A_i$ acting on $V$ and complex coefficients $c_i$ such that $\sum_i c_i =
1$. (Almost all trace-preserving operators are of this form). The hermiticity
condition implies that $A_i$ are unitary, while the positivity condition that $c_i$
are non-negative. So $Y$ is a convex combination of unitary automorphisms
\begin{equation}
Y(O) = \sum_i c_i U_i O U_i^{\dagger}, \label{convex}
\end{equation}
for a family of unitary operators $U_i$.

\subsubsection{Time symmetries}
Now, there are certain properties concerning the temporal properties of the quantum
process that can be written as symmetries of the propagator $\upsilon$.
\\ \\
{\em Time homogeneity:} A quantum Markov process
 is called {\em time homogeneous} if
$\upsilon(z,z';t|z_0,z'_0;s) = \upsilon(z,z';t|z_0,z'_0;0)$.  This implies that the
information of the propagator is contained in the one-parameter family of kernels
$\upsilon_t(z,z'|z_0,z_0')$. Note, that in the physics literature, it is often the
time-homogeneous Markov processes that are referred to as Markov processes.

If we translate this condition in terms of the operator  $Y$ on $V \otimes \bar{V}$,
it implies that $Y_t$ forms an one-parameter group of trace-preserving
transformations.
\\ \\
More important is the following condition \\ \\
{\em Time reversibility:} A quantum Markov process in an  interval $[t_0,t_f]$ is
called {\em time-reversible} if
\begin{equation}
\upsilon(z,z'; (t_f -t_0)-t| z_0,z'_0; (t_f - t_0)-s)  =
\bar{\upsilon}(z,z';t|z_0,z'_0;s) \label{timerev}
\end{equation}
This condition implies that one can interpret the two  entries of the decoherence
functional as a path going forward in time and one going backwards, implying that its
evaluation takes place in a closed time-path.

For this purpose one could write more naturally the  condition of {\em strong time
reversibility}
\begin{eqnarray}
\upsilon_{n,m}(z_1,t_1; \ldots; z_n,t_n|z'_1,t'_1; \ldots ; z'_m,t'_m) = \nonumber \\
\bar{\upsilon}^{n,m}(z'_1,t'_1; \ldots ; z'_m,t'_m| z_1,t_1; \ldots ; z_n, t_n)
\label{stimerev}
\end{eqnarray}
This is, however, too strong: all physical decoherence functionals {\em break } time
reversibility by virtue of containing information about an initial condition at a
single moment of time. The only way to satisfy this is with a decoherence functional
containing both initial and final conditions, and such that these conditions are in
some sense equivalent \footnote{  In Hilbert space quantum theory one can insert  a
final-time density matrix $\hat{\rho}_f$ in equation (\ref{decfun}). The theory is
then strongly time reversible if $\hat{\rho}_f = e^{- i \hat{H}(t_f -t_0)}
\hat{\rho}_0 e^{i \hat{H}(t_f - t_0)}$. Such theories are not employed as all
preparations of the system (by which we condition and define the initial state) take
place in the past of the measuring procedure. This construction is not nonsensical,
but (at least seemingly)  void of any operational content. If, however, we decide to
extend the scope of quantum theory at the realm of cosmology, such conditions cannot
be excluded and they may hold an appeal because of their strong symmetry. }. For this
reason, the condition (\ref{timerev}) which singles out the dynamical notion of
time-reversibility is preferred.  We should note, a separation between dynamics and
initial condition makes sense only when the Markov property is satisfied.

 The condition of time-reversibility implies that the
 operator $Y$ on $V \otimes \hat{V}$ is
{\em unitary}, i.e. that
\begin{equation}
Tr(A^{\dagger}Y(B)) = Tr(Y^{-1}(A)^{\dagger}B) \label{timerevop}
\end{equation}

Now, if $A$ and $B$ are the one-dimensional projector $P$ this condition implies that
\begin{equation}
Tr(Y(P)^{\dagger} Y(P)) = Tr(P^2) = TrP = 1.
\end{equation}
This implies that $Y$ preserves pure density matrix and it is a theorem \cite{Davies}
that it should be of the form
\begin{equation}
Y(O) = UOU^{\dagger},
\end{equation}
for some unitary operator $U$. One way to see this without referring to the lengthy
proof   is to notice that operators of the form (\ref{convex}) satisfy property
(\ref{timerevop}) if and only if $c_i = 0$ for all but one values of $i$. This does
not constitute a proof, however, because not all operators that satisfy the
hermiticity, trace-preservation and positivity properties can be written in the form
(\ref{convex}).

In any case, time reversibility implies that  $Y = U \otimes U^{\dagger}$. This means
that
 the propagator $\upsilon$ factorises
\begin{equation}
\upsilon(z,z';t |z_0,z'_0 ;s) = \chi(z,t|z_0,s) \bar{\chi}(z',t'|z'_0,s)
\end{equation}
for a kernel  $\chi$ on $\Gamma_t \times \Gamma_s$, that corresponds to the unitary
operator $Y$. This clearly satisfies
\begin{equation}
\chi(z,t|z_0,s) = \bar{\chi}(z_0,s|z,t).
\end{equation}

This kernel   will be eventually interpreted as the  wave function propagator in the
coherent state basis. If, in addition, the process is time homogeneous one can write
\begin{equation}
\chi(z,t|z_0,s) = \chi_{t-s} (z|z_0),
\end{equation}
in terms of an one-parameter family of kernels $\chi_t$ on $\Gamma \times \Gamma$.

In a nutshell, a quantum process that satisfies the Markov property is completely
characterised by the knowledge of a  single time kernel $\upsilon_0$ and  of the
propagator $\upsilon$. If it is time-symmetric the propagator factorises and is
determined in terms of a single kernel $\chi$ on $\Gamma_t \times \Gamma_s$.

\subsection{Deriving quantum theory}
We will now show how the Hilbert space of standard quantum theory naturally arises in
the study of quantum processes. First let us emphasise once more, that the Hilbert
space $V$ we employed in the previous section {\em is not} the  Hilbert space of
standard quantum theory. The latter is actually a subspace of $V$ and it is, in
general, non-trivial to determine how this subspace would be selected.

\subsubsection{The physical Hilbert space and the Hamiltonian}
 Now, let us suppose that we have a quantum process
$(\Omega, \Phi, F^a_t)$, where $\Omega $ is a subset of $\times_t \Gamma_t$, $\Gamma$
is a manifold (to be identified with the classical phase space) and $F^a_t$ correspond
to coordinates on $\Gamma$. Assume further that this process satisfies the Markov
property.

As we showed previously the Markov property implies that the process is uniquely
determined by the knowledge of a kernel $\upsilon_0(z_0,z'_0)$ at some reference time
$t_0$ and the propagating kernel $\upsilon(z,z';t|z_0,z_0';s)$. If we also assume
that the process is {\em time reversible}, then one needs simply to specify a
propagating kernel of the type $\chi(z,t|z_0,s)$.

Let us now study the kinematical process associated with this quantum process by
considering the decoherence functional generated by $\upsilon_0$ and a propagating
kernel $\psi_t(z|z') = \lim_{s \rightarrow t} \chi(z,t|z',s)$. We need to {\em assume
that $\psi_t$ is a continuous and perhaps sufficiently smooth function on $\Gamma$
}\footnote{Note if we study the standard quantum process on {\em configuration space}
the kernel $\psi_t$ defining the kinematical process is a $\delta$-function, which
cannot straightforwardly define a quantum process. This is unlike the case of phase
space processes we consider here.}.

The quantum Chapman-Kolmogorov  equation is valid for the kinematical process also so
\begin{equation}
\psi_t(z|z') = \int dz'' \psi_t(z|z'') \psi_t(z'',z')
\end{equation}
This is, however, the defining equation for a projection operator $E_t$ in the space
$V = {\cal L}^2(\Omega,dz)$. The range of this projector is a Hilbert space $H_t$,
corresponding to the instant $t$.

If the kinematical process is {\em time-homogeneous}, i.e. $\psi_t$ does not depend on
the time $t$, then all $H_t$ are naturally identified with a reference Hilbert space
$H$, which is the Hilbert space of standard quantum theory.

Let us now  consider the kernel $\chi_t(z|z')$. The quantum Chapman-Kolmogorov
equation implies that
\begin{eqnarray}
\chi_t(z|z') = \int dz'' \psi(z|z'') \chi_t(z''|z') = \int dz'' \chi_t(z|z'')
\psi(z''|z')
\end{eqnarray}

In terms of operators on $V$, this reads
\begin{equation}
U_t = E U_t = U_t E,
\end{equation}
implying that $U_t = E U_t E$. The kernel $\chi_t$ then defines an one-parameter
family of unitary operators on $H$; having assumed continuity at $t = 0$, we can
employ Stone's theorem to determine that it would be of the form $e^{-i \hat{H}t}$ in
terms of a self-adjoint operator on $H$.

We need to repeat again that the finiteness and continuity of $\chi_t$ is {\em
necessary} if we are to obtain the Hilbert space $H$ and the corresponding
Hamiltonian. In absence of this condition, there is no way to pass from the unphysical
Hilbert space $V$, to the Hilbert space $H$.

Some further remarks can be made at this point. Consider the kernel $\rho_0$
containing the information about the initial state. It corresponds to an operator
$\hat{\rho}_0$ on $V$. Since the restriction of the decoherence functional at $t_0$
still has to satisfy the axioms of the decoherence functional, we see that
$\hat{\rho}_0$ is a density matrix on $V$. Now, any integration over $z_i$ or $z'_i$
in equation (\ref{factorisation}) (for the determination of a correlation function)
will essentially amount to squeezing $\hat{\rho}_0$ between two projectors $E$ as $E
\hat{\rho}_0 E$. This means that effectively the initial condition will be encoded in
a density matrix on $H$, rather than $V$, which is the standard result.

\subsubsection{The coherent state picture}

Now, on $V$ one can write the coherent states $\Psi_z$ as the functions $\Psi_z(z') =
\chi(z'|z)$. The whole Hilbert space $H$ can be obtained from finite linear
combinations of these coherent states.
 Clearly then, $\chi_t(z,z') = \langle z'|e^{-i \hat{H}t}|z \rangle$
 and the distribution functions $\upsilon^{n,m}$ will read
\begin{eqnarray}
\upsilon^{n,m}_{z_0}(z_1,t_1; z_2,t_2; \ldots ; z_n, t_n|z'_1,t'_1;
z'_2, t'_2; \ldots ; z'_m,t'_m) =  \nonumber \\
\langle z'_m|e^{-i \hat{H}(t'_m - t_n)}| z_n \rangle \langle z_n| e^{-i
\hat{H}(t_{n}-t_{n-1})}| z_{n-1} \rangle \ldots
\nonumber \\
\times \langle z_2|e^{-i \hat{H}(t_2-t_1)}| z_1 \rangle \langle z_1|e^{-i
\hat{H}(t_1-t_0)}| z_0 \rangle \langle z_0|\hat{\rho}_0|z'_0 \rangle\nonumber \\
\times \langle z'_0| e^{i \hat{H}(t'_1-t_0)}| z'_1 \rangle \langle
z'_1|e^{i\hat{H}(t'_2-t'_1)} | z'_2 \rangle \ldots \langle z'_{n-1}|e^{i
\hat{H}(t'_n-t'_{n-1})}| z'_n \rangle ,
\end{eqnarray}
which is the standard quantum mechanical result.

Furthermore if we consider  functions $f^a$ on $\Gamma$, we can write the correlation
functions for the process corresponding to $F_t^a$
\begin{eqnarray}
G^{n,m}(a_1,t_1; \ldots; a_n,t_n| a'_1,t'_1; \ldots ; a'_m,t'_m)
= \nonumber \\
Tr \left( \hat{A}^{a_n}(t_n) \ldots \hat{A}^{a_1}(t_1) \hat{\rho}_0
\hat{A}^{a'_1}(t'_1) \ldots \hat{A}^{a'_m}(t'_m) \right) .
\end{eqnarray}
In this equation we have assume temporal ordering as $t_1 < t_2 < \ldots < t_n$ and
$t'_1 < t'_2 < \ldots < t'_m$. The operators $\hat{A}(t) = e^{i \hat{H}(t-t_0)}
\hat{A} e^{-i\hat{H} (t-t_0)}$ are the Heisenberg picture operators and
\begin{equation}
\hat{A}^a = \int dz f^a(z) |z \rangle \langle z |
\end{equation}

Furthermore, if we have a pair of discrete histories $\alpha = (C_1,t_1; \ldots ;
C_n,t_n)$ and $\beta = (C'_1,t'_1; \ldots ; C'_m,t'_m)$, where the $C_i$ and $C'_i$
are measurable subsets of $\Gamma$, we will have the value of the decoherence
functional
\begin{equation}
\Phi(\alpha, \beta) = Tr \left( \hat{C}_n(t_n) \ldots \hat{C}_1(t_1) \hat{\rho}_0
\hat{C}_1'(t'_1) \ldots \hat{C}_m'(t'_m) \right),
\end{equation}
where here $\hat{C} = \int_C dz |z \rangle \langle z|$ are positive operators that
represent the phase space cell $C$. They have been used in \cite{Omn4} in the
discussion of the classical limit. Again $\hat{C}(t)$ denotes the Heisenberg-picture
operator.

This expression for the decoherence functional is identical with the standard
expression (\ref{decfun}); the only difference is that in our case events {\em are not
represented by projectors}. Events correspond to measurable subsets of $\Gamma$, which
form a Boolean algebra and as such they are represented by a particular class of
positive operators.

The assumption of time-reversibility is not essential in obtaining standard results of
quantum theory. One can construct a kinematical process corresponding to
$\upsilon(z,z';t|z_0,z_0';s)$. In the case of a time-homogeneous process the
corresponding kernel defines a projection operator into a closed linear subspace $U$
of $V \otimes \bar{V}$, which satisfies the hermiticity property. Similarly the
dynamics correspond to an operator that is projected on $U$; effectively $Y_t$
corresponds to a {\em bistochastic map}
 characterising an {\em open quantum system} \cite{Str96}.

\subsubsection{A summary}
Let us recapitulate the results, so far, in this section.
If we assume we have a {\em quantum process} that \\ \\
1. Satisfies the Markov property,
\\
2. Its propagator $\upsilon$ is a continuous function of the time variable as well as
the points $z \in \Gamma$,
\\ \\
then we recover standard quantum theory for open systems (of the Markov type).
 \\ \\
3. If in addition we demand that  the quantum process is time-reversible \\ \\
 we can define a Hilbert
space {\em for each moment of time}, but it is not necessary that this can be reduced
to the standard case of a single Hilbert space describing the whole system. We have
analysed this feature of quantum mechanical histories (with reference to quantum field
theory in curved spacetime) in \cite{An00a}.
\\ \\
4. If we  also demand that  the quantum process is time-homogeneous, \\ \\
we can write a reference Hilbert space describing the system at all times, verify that
the dynamics are given by a Hamiltonian operator and obtain standard quantum theory
with all its predictions but {\em one}.

Namely we get the correct correlation functions for the observables of quantum theory,
but our events do not correspond to {\em projection operators}. They correspond to a
larger class of positive, bounded operators \footnote{One can easily check that in
terms of the supremum norm $||\hat{C}|| \leq 1$.}.

In other words the theory of quantum processes gives an equivalent description of the
one afforded to quantum theory (as far as statistical predictions are concerned), {\em
while  it does not abandon the classical ``logic'' for the description of events}.

\subsubsection{Some remarks}
At this point we need to make a number of important remarks
\\ \\
1. The fact that the physical Hilbert space is not identical with the mathematically
natural one $V$, is due to the fact that $\psi(z|z')$ is a genuine function over
$\Gamma \times \Gamma$. Thus it can be used to define a projector. If this were not
the case and $\psi$ were a delta function the corresponding projector would be unity,
and $V$ would be identical with $H$. Such is the case in classical probability theory;
in this case {\em the kinematic process is completely trivial}. Hence one might say
that the non-triviality of the kinematical process is the reason, why the physical
Hilbert space is a subspace of $V$.
\\ \\
2. In equation (\ref{Psymbol1}) we chose to employ the $P$-symbol as giving the
correspondence between quantum mechanical operators and phase space functions. This
choice was at that point arbitrary; there is a continuous infinity of possible symbols
one could have employed for the correspondence, the $P$-symbols being just one
possibility. However, in our analysis of section 5, the $P$-symbol arises {\em
naturally} as giving the correspondence between functions and operators. In our
argumentation, we started with a quantum process on phase space, hence the functions
$f^a(z)$ are the fundamental variables and the operators are secondary  objects; given
the way operators appear as a convenient description, equation (\ref{Psymbol1}) for
the $P$-symbol is  the only one that physically makes sense.
\\ \\
3. In defining the quantum Markov property we have imposed a condition on {\em
conditional pairs}, which is the mathematically natural thing to do. However, the
incorporation of measurement {\em outcomes} in the decoherence functional takes place
with respect to {\em conditional correlation}, as this is equivalent to the
``reduction of the wave packet'', and this object is too unwieldy to yield a useful
definition of a Markov process. This distinction is in contrast to classical
probability theory, where the conditional expectation is used for {\em both} defining
the Markov property and incorporating experimental results. This difference points to
a fundamental difference of the notion of events between classical and quantum theory
(whenever we attempt to have a realist description of the theory). This is not an
issue that we can further elaborate at this point.
\\ \\
4. Standard quantum mechanics is recovered only  when we have a Markov process. This
is natural as the notion of state makes little sense otherwise; {\em a state} is a
description of the system at a moment of time. It  contains information about
possible  measurements at this instant,  But it should also  contain information that
will allow us to determine (or predict) the state at subsequent moments of time
(assuming we know the dynamical law). In absence of the Markov property, no
information at a moment of time suffices to determine the future development of the
system.

\subsection{Phase space structures in quantum processes}
We have seen that  a time-reversible, Markov quantum process can be determined by the
propagator $\chi_t(z|z')$, $z, z' \in \Gamma$. The question then arises, how such an
object can be constructed solely from geometric structures on the symplectic manifold
$\Gamma$. One answer comes from the coherent state quantisation programme developed by
Klauder \cite{Kla88}.

The idea is that the propagator in the coherent state basis can be obtained as a
suitably regularised path integral. The regularisation takes place with respect to a
{\em homogeneous'} Riemannian metric $g$ on $\Gamma$. This metric can be used to
define a conditioned Wiener process on $\Gamma$, with fixed points at initial and
final moments $z(0) = z_0$ and $z(t) = z_f$. The corresponding reads formally
\begin{equation}
d \mu^{\nu}_{z_0,z_f} [z(\cdot)] = \exp \left( - \frac{1}{2 \nu} \int_0^t
g(\dot{z},\dot{z}) ds \right) Dz(\cdot).
\end{equation}

Then it turns out that one can write
\begin{equation}
\chi_t(z_f|z_0) = \lim_{\nu \rightarrow \infty} 2 \pi e^{\nu t/2} \int e^{i
(\int_{z(\cdot)} A  - \int_0^t h(z(s)) ds)} d \mu^{\nu}_{z_0,z_f}[z(\cdot)].
\end{equation}
In this expression $A$ is a connection one-form such that $dA = \omega$, with $\omega
$ the symplectic form on $\Gamma$. Such an one-form arises from the procedure of
geometric quantisation, i.e. identifying a $U(1)$ bundle over $\Gamma$, with a
connection whose curvature projects to $\omega$. The Hamiltonian $h(z)$ is essentially
the $P$-symbol of the quantum Hamiltonian operator.

As Klauder has emphasised the whole procedure is fully geometrical; if we want to
construct the projection operator $E$ that defines the physical Hilbert  space we need
to specify the connection $A$ and the Riemannian metric, since
\begin{equation}
\psi(z_f|z_0) =  \lim_{\nu \rightarrow \infty} 2 \pi e^{\nu t/2} \int e^{i
\int_{z(\cdot)}A } d \mu^{\nu}_{z_0,z_f}[z(\cdot)].
\end{equation}
In order to specify the dynamical component we need to specify a scalar function on
$\Gamma$ that will act as the Hamiltonian.

\section{Interpretational issues}

As we explained, the difference between quantum processes  and standard quantum theory
lies only in the determination of which object correspond to sharp events. Quantum
mechanics admits projection operators, while the theory of quantum processes admits
phase space cells. These are represented by a positive operator-valued-measure $C
\rightarrow \hat{C} = \int_C dz |z \rangle \langle z|$, for any measurable subset $C$
of $\Gamma$.

The question then arises, which of the basic principles of quantum theory is (are )
violated by this change and whether this violation has empirical consequences.
\subsection{Comparing with standard quantum theory}
Many textbooks employ the following basic principles (or some variations of them) as
axioms, out of which the basic  structure of quantum theory is derived.
\\ \\
D1. {\em States:} A state of the system at a moment of time is represented
by a vector on a Hilbert space $H$, or a density matrix on $H$. \\
D2. {\em Observables:} Observables are represented by self-adjoint  operators on $H$. \\
D3. {\em Properties:} The possible values for an observable correspond  to the points
of the spectrum of the corresponding operator. As a corollary a proposition about
possible values of an observable is represented by a projection operator.
\\
D4. {\em Probabilities:} If $\hat{P}$ is a projection operator
 representing a property,  then the probability that this is
 true in a state $\hat{\rho}$ equals $Tr(\hat{\rho} \hat{P})$. \\
D5. {\em Combination of subsystems:} If $H_1$ and $H_2$ are Hilbert spaces
characterising independent systems, the combined system is
characterised by $H_1 \otimes H_2$. \\
D6. {\em Conditioning:} If an experiment verifies a property corresponding
 to a projection operator $\hat{P}$, then we encode this information by
 transforming the state as
 $\hat{\rho} \rightarrow \hat{P} \hat{\rho} \hat{P} / Tr(\hat{\rho} \hat{P})$. \\
D7. {Time evolution:} If the system is left isolated, its state  evolves under the
action of an one-parameter group of unitary transformations.
\\ \\
In order to facilitate comparison, let us gather here the basic principles  of the
theory of quantum processes:
\\ \\
E1.{\em Observables and events:} A physical system is characterised by
 a history space $\Omega$, that is a suitable subset of a Cartesian product
 $\times_{t} \Gamma_t$, for some manifold $\Gamma$. Events correspond to Borel
 subsets of $\Omega$ and observables to functions on $\Omega$. \\
E2. {\em States:} The decoherence functional is a bilinear functional  on $\Omega$
satisfying properties C1-C6.
\\
E3. {\em Probabilities and phases:} If $A$ and $B$ are two histories (subsets of
$\Omega$) then $\Phi(A,A)$ gives the probability that  $A$ is true and $\Phi(A,B)$
gives the relative Pancharatnam phase, measured
 in the way described in section 2.4. \\
E4. {\em Combination of subsystems:} If $\Omega_1$ and $\Omega_2$ are describe two
physical systems, then $\Omega_1 \times \Omega_2$ describes the combined system.  \\
E5. {\em Conditioning:} One conditions with respect to  a subalgebra using the
conditional pair and incorporates information about previous experiments through the
conditional correlation (see section 3.4.2).
\\ \\
Let us now recall the results of section 5. Principles D1, D2 arise in  quantum
processes that satisfy the Markov, time-reversibility and time-homogeneity properties.
Principle D3 is the one where the two theories disagree and we shall examine it in
more detail in a moment. Principle D4 is equivalent to E3, when the decoherence
functional is restricted to a moment of time.

Concerning the combination of subsystems, let us note the following.  If $\Gamma =
\Gamma_1 \otimes \Gamma_2$ is the phase space at a moment of time in a Markov process
the space $V = {\cal L}^2(\Gamma, dz_1 dz_2)$ naturally appears and is isomorphic to
${\cal L}^2(\Gamma_1, dz_1) \otimes {\cal L}^2(\Gamma_2, dz_2) = V_1 \otimes V_2$. Now
if the two processes are independent, the decoherence functional will be $\Phi_1
\otimes \Phi_2$, which in a Markov process implies that both the initial state
$\upsilon_0$ and the propagator $\chi$ factorises as
\begin{equation}
\chi_t(z_1,z_2|z'_1,z'_2) = \chi_{1t}(z_1|z'_1) \chi_{2t}(z_2|z'_2).
\end{equation}
Clearly the projector $E$ that corresponds to the kernel $\psi = \lim_{t \rightarrow
0} \chi_t$ also factorises and the resulting physical Hilbert space can be written as
$H = H_1 \otimes H_2$. Note that if the two subsystems are not independent, the
physical Hilbert space is {\em still} a subspace of $V_1 \otimes V_2$.

As we discussed in 3.4.2 the notion of the conditional correlation recovers  the
standard results for the ``reduction of the wave packet `` rule D6. And finally   a
time-reversible, Markov quantum process can be described by an one parameter family of
unitary operators. If the process   is also time-homogeneous, then  this family of
operators is an one-parameter group.

\subsection{Do we measure operator's eigenvalues?}
So the only difference between the theory of quantum processes and quantum mechanics
is the principle D3: in a quantum process the spectrum of an operator is simply not
relevant to the values of the corresponding observable, because at the fundamental
level observables are functions on the history space $\Omega$. Clearly there is little
difference as far as observables with continuous spectrum are concerned (position,
momentum etc). The difference lies, of course, in the case of observables with
discrete spectrum.

The case of discrete spectrum is, in fact, what has given quantum phenomena their
name, as it is this through the discrete spectrum of operators that the paradigmatic
quantum behaviour is manifested: historically it was the black body radiation, the
photoelectric effect  and the Bohr's atom transitions that put discreteness as a basic
feature of the new mechanics. For this reason postulate D3 was highlighted  in all
early work of quantum theory: it provided  a simple  solution to the problems that had
faced a generation of physicists. Later mathematical development -namely the spectral
theorem - offered this postulate the additional justification of mathematical
elegance.

It would seem that this is one of the most solid postulates of quantum  mechanics, the
last one to be taken away from any possible modification of the theory. After all it
provides the solution to the physical puzzles that led to quantum  mechanics. However,
as we are going to argue it is the postulate of quantum theory that is {\em least}
justified empirically, when taken by itself.

To see this we shall consider the case of atom spectroscopy, which has been
 historically the main arena justifying the postulate D3.
 When we study  the
electromagnetic radiation emitted from atoms, we see that the intensity of the
electromagnetic field has peaks in particular discrete values of the frequency. Then
assuming energy conservation, the photons are viewed as arising from  a transition
between two "states" of an atom , each of which is characterised by a sharp value of
energy. The fact that we measure  a  number of sharp peaks rather than a smoother
distribution of field intensity plotted versus frequency, leads to the conclusion that
the possible values of atom's energy are discrete. If we assume that this experiment
measures the atom's energy, then this takes discrete values, something that is
naturally explained in terms of postulate D3: in any individual measurement only
points of the spectrum of the operator are obtained.

We believe that this is a fair summary of the argument that leads to the acceptance of
the postulate D3 in the particular context. We shall now see, that the conclusions of
the argument is by no means necessary. Let us first make the too obvious remark, that
the  intensity peaks have finite width and are not sharp. The width is due not only to
experimental errors, but comes fundamentally from the time-energy uncertainty
relation. Hence, it is only in an idealisation that the atom's energy values are
discrete.

However, the most important argument is that the description in terms of atom
transitions {\em is semiclassical rather than quantum}. What we measure in
spectroscopy is  the energy/frequency of the electromagnetic field. We typically
assume that the emitted photons are incoherent (both in the classical and the quantum
sense), so that the emitted electromagnetic field can be considered as an ensemble of
photons. Then, we can idealise the experiments by means filters that allow only very
narrow frequency (energy) range to pass and measure the intensities. The whole
experiment is then fully described by energy measurements of the photons. One can give
an equivalent description in terms of the electromagnetic fields. So the actual
observables that correspond to the set-up of the experiment is photon energies or
fluxes, {\em not atomic energies}. And these energies can be described by continuous
variables in either quantum theory or in the quantum process description.

The attribution of discrete energy values to the atom comes from a semiclassical {\em
picture} of the atom/field interaction; it involves a mixture of old quantum theory
concepts (orbitals, transitions), with the framework of mature quantum theory. This
picture is helpful for calculations, it provides an intuitive picture of the
interaction, but it is not fundamentally quantum mechanical. A precise treatment ought
to consider the combined system field-atom, interacting perhaps through QED and then
consider energy measurements of the electromagnetic field at particular spatial
locations. In such a description all information about the process (including the
atom's eigenvalues) would be found in the correlation functions of the electromagnetic
field: {\em but these are predicted by quantum processes in full agreement with
standard quantum theory}.

What we imply by this argument, is that historically the discrete values of
observables  refer to the spectrum of the Hamiltonian, rather than any arbitrary
observable. The information about its eigenvalues is fully contained in the
correlation functions: once these are provided, we can read off any discretised
behaviour. In other words, {\em the discrete behaviour in quantum theory is not
fundamental, but arises due to particular forms of the dynamics.} This is true even
for spin systems: the "discrete" spin values are always measured in conjunction with
its coupling to some magnetic field.

Discrete behaviour arising from dynamics is not something strange or new. The
stochastic mechanics developed by Nelson \cite{Nel85} is a framework trying to
describe quantum theory in term of stochastic processes on configuration space: as in
our case, stochastic mechanics considers fundamentally continuous observables and
generates discrete structures dynamically. Stochastic mechanics ultimately fails as an
alternative theory to quantum mechanics,  since it cannot account for Bell's theorem,
however it has provided many examples by which stochastic process simulate discrete
behaviour, actually reproducing quantum mechanical phenomenology. The reader is
referred to \cite{Nel85, GORW78} for elaboration of  arguments similar to the ones we
presented here.

We conclude our argument by noticing, that of all axioms of Kopenhagen quantum theory,
it is only D3 that makes implicit reference to properties of individual quantum
systems (namely in any individual system we measure certain particular values for an
observable). If we take the stance that Kopenhagen is a statistical theory of
ensembles, axiom D3 might be viewed as the odd one among all quantum mechanical
postulates.

\subsection{Preparation and conditioning}
Another point we want to address is the notion of preparation. The controlled
preparation of quantum  state is the first important part of any measurement scheme.
There are various procedures that are followed in actual experiments: these procedures
have to be repeatable and aim to provide an accurate description of the system in
terms of a well specified Hilbert space vector or density matrix.

In the theory of quantum processes one represents  the effects of preparation by the
notion of conditioning. We need first to assume a basic decoherence functional for the
system: this is not the effective decoherence functional we employ for the
measurements, it is just a decoherence functional representing the systems under study
before even preparation. Its actual form will probably play little role in the set-up
of the measurement process. Thus it can be a decoherence functional of the standard
form, with initial density matrix corresponding to total ignorance; or it can be of a
 simple factorised form
\begin{equation}
\Phi(A,B) = \sigma(A) \bar{\sigma}(B).
\end{equation}

Whatever the initial form might be, as soon as the experimentalist starts interfering
with the system, he will not employ the initial decoherence functional in his
calculations. Rather he has to transform the initial decoherence functional through
the incorporation of all information corresponding to the preparation procedure. If,
for instance, the experimentalist  passes a beam of particles through a filter so that
only ones with a given property will be further studied, the decoherence functional
has to transform according to the standard rules.

The preparation of the system is then described by successive conditionings of the
decoherence functional, plus evolution that needs be neither unitary (systems are
often coupled to external reservoirs so that they relax in their ground state), or
even Markovian (the experimentalist can choose to perform or not a given operation
according to the outcome of a past preparation procedure).

In principle, one can describe the whole sequence of preparation procedures, either as
conditioning of the decoherence functional, or introduction of "unusual" dynamics at
different times $t$. However, physical systems seem to be described (when isolated) by
Markovian dynamics, so at the moment $t = 0$, where the last step of preparation has
been concluded, the whole information of the preparation  can be encoded in its
restriction at $t=0$ (in other words a density matrix) and its evolution operator: the
process is then Markovian for all times $t \geq 0$; and naturally it is not Markovian
if the whole of the time axis is taken as describing it.

We should keep in mind, that if the physical systems were not described by a Markov
process, the description in terms of an initial state would be meaningless, since any
step in the preparation procedure might affect (unexpectedly) later evolution of the
system.

The other point to note, is that typical preparation procedures aim to force the
quantum system to forget its past or to make the past   irrelevant for the
experimental procedure: the coupling to a reservoir, for instance, brings all systems
to their initial state. So it seems fair to assume that the initial form of the
decoherence functional before conditioning, will be largely irrelevant to the
determination of the density matrix that arises out of the preparation. In this
context, the question whether there is some decoherence functional describing the
system outside operational situations can have no answer; at least not until we have
an acceptable quantum theory for individual systems.

\subsection{Further remarks}

As far as the interpretation of quantum theory is concerned, the theory of quantum
processes has a feature, which we think is a distinct advantage over standard quantum
theory. Its "logic" is classical. Even though we have cast the theory in a Kopenhagen
form, there is no problem in talking about properties of individual systems outside a
measurement context (there is no problem of definability), as long as we remember that
all our predictions refer to ensembles. This theory completely sidesteps the obstacles
of theorems such as the one of Kochen and Specker for the existence of {\em
uncontextual} realist theories. A full discussion about these issues is to be found in
\cite{An01a}.

In the same reference it is also discussed, whether the difference between quantum
processes and standard quantum theory in terms of the notion of event can be
experimentally determined. The two theories differ in their statistical correlation
functions (\ref{statcor}), while they have the same quantum mechanical correlation
functions. The answer tends to be negative, even though the possibility of devising an
experiment that distinguishes them cannot be ruled out. The reason is -in a sentence-
that in realistic experiments the actual time evolution (due also to the coupling of
the system with the measuring device) tends to blur the distinction, because quantum
mechanical time evolution does not preserve the characteristic functions on the phase
space. The reader is again referred to \cite{An01a} for a fuller analysis.

\section{Conclusions}
In this paper we described a theoretical framework for quantum theory, which is
distinct from standard quantum theory. It is a
development of the  following basic ideas \\ \\
i. Quantum theory provides information not only about probabilities but also of
complex phases relative to different histories: these phases are measurable in the
same way as the Pancharatnam phase. The information about both phases and
probabilities is encoded in the decoherence functional, which also contains all
information about the system's correlation functions.
\\
ii. Putting complex phases as primitive ingrdients of theory leads to a description of
a system in terms of non-Kolmogorov probability. All constraints coming from non-go
theorems do not hold here, hence we can write a theory with commutative observables
(i.e. "hidden variables") that can reproduce the results of standard quantum theory.
\\
iii. The classical phase space of the system contains enough information, to fully
incorporate all quantum mechanical behaviour. Using the formal analogy with stochastic
processes, we can set up a theory of {\em quantum processes} on phase space: the basic
object will be now the decoherence functional rather than the probability measure.
\\ \\
We then saw, how this construction can be developed and recover standard quantum
theory,  by demanding the satisfaction of the Markov property and that the kinematic
process is not trivial.

What do we think are the basic achievements of this paper?

We showed, that there is the possibility of a different axiomatic framework for
quantum theory, that does not necessitate the introduction of non-commutative objects
as fundamental. The distinctive  quantum mechanical behaviour is contained in an
object that plays the role of the density matrix for histories: the decoherence
functional. In our construction, there is no need to consider that quantum theory
necessitates a radical change in the concept of events (as  quantum logic is often
taken to imply). Our logic remains Boolean. We can say that we have a theory of hidden
variables; they are non-deterministic, of course, but they are fully compatible with
realism.

We could then come forward and claim that the theory of quantum processes provides an
alternative description of quantum phenomena, from which the Hilbert space description
arises as a convenient special case. Our arguments were completely general and we did
not specialise on any particular system: it suffices that one can write an effective
phase space description for the quantum system. Our result are valid also for bosonic
fields, but they are not readily employed for spinor fields. The reason is that, as
yet we lack a phase space description of spinor fields (through coherent stace) that
corresponds to a theory with local Hamiltonian (the existing fermionic coherent states
\cite{Kla59}  have non-local dynamics). We are trying to address the issue by
constructing coherent states for the fermion fields, starting from coherent states of
the corresponding relativistic particles. With the same motivation
 we have provided a relevant discussion for the
spin-statistics theorem \cite{An01d}. If we succeed in this endeavour, we shall be in
a position to claim that a description of quantum processes can supplement standard
quantum theory in {\em all physical systems} accessible to experiment.

One could then raise the issue of naturality: how natural is the construction of
quantum processes as compared to standard quantum theory? This is a difficult question
to answer: by what criteria is a notion of a functional containing phase and
probability information more (or less) natural than the notion that observables are
actually non-commutative objects? What is natural is often an issue of universality,
but sometimes it is  simply a matter  of habit.

Our eventual aim is, however, different from either the interpretational issue or the
rather interesting mathematical constructions a theory of quantum processes can lead
to. After all, we do not provide different domain of  applicability  from standard
quantum theory. Now, if we accept that the notion of events in quantum theory is not
different from the one of classical theory, there is a natural question to ask. We
have a successful statistical theory for physical systems that gives non-additive
probabilities, which are related to some  relative $U(1)$ phases of unknown origin;
can we explain this statistical behaviour in terms of properties of the individual
system? To lead our investigations towards the construction of such  a theory, is the
final aim of this paper.

\end{document}